\documentclass[aps,pre,twocolumn,groupedaddress,showpacs]{revtex4-1}

\usepackage{mathtools}
\usepackage{amsmath}
\usepackage{amssymb}
\usepackage{bm}
\usepackage{graphicx}
\usepackage{xcolor}
\usepackage{epstopdf}
\usepackage{subfig}
\usepackage{subfloat}

\newcommand{\lx} {\left}
\newcommand{\rx} {\right}
\newcommand{\eps} {\epsilon}

\newcommand{\ave}[1] {\lx\langle #1 \rx\rangle}
\newcommand{\be} {\begin{equation*}}
\newcommand{\ee} {\end{equation*}}
\newcommand{\bea} {\begin{eqnarray*}}
\newcommand{\eea} {\end{eqnarray*}}

\newcommand{\cD} {\mathcal{D}}

\newcommand{\cP} {\mathcal{P}}
\newcommand{\cN} {\mathcal{N}}

\newcommand{\cT} {\mathcal{T}}

\newcommand{\xif} {\widetilde{\xi}}
\newcommand{\xifo} {\xif(\om)}

\newcommand{\om} {\omega}
\newcommand{\xf} {\widetilde{x}}
\newcommand{\xfo} {\xf(\om)}

\newcommand{\cL} {\mathcal{L}}
\newcommand{\cLf} {\widetilde{\cL}}
\newcommand{\cLfo} {\cLf(\om)}
\newcommand{\cR} {\mathcal{R}}
\newcommand{\cRf} {\widetilde{\cR}}
\newcommand{\cRfo} {\cRf(\om)}

\newcommand{\cG} {\mathcal{G}}
\newcommand{\cGf} {\widetilde{\cG}}
\newcommand{\cGfo} {\cGf(\om)}

\newcommand{\cA} {\mathcal{A}}
\newcommand{\cAf} {\widetilde{\cA}}
\newcommand{\cAfo} {\cAf(\om)}

\newcommand{\cB} {\mathcal{B}}
\newcommand{\cBf} {\widetilde{\cB}}
\newcommand{\cBfo} {\cBf(\om)}

\newcommand{\cSgf} {\widetilde{\Sigma}}
\newcommand{\cSgfo} {\cSgf(\om)}

\newcommand{\cC} {\mathcal{C}}
\newcommand{\cCf} {\widetilde{\cC}}
\newcommand{\cCfo} {\cCf(\om)}

\newcommand {\Pab} {\mathbf{G}}
\newcommand {\pa} {\bm{\Pi}}
\newcommand {\one} {\mathbf{1}}
\newcommand {\sa} {a}
\newcommand {\bsa} {\bm{a}}
\newcommand {\obsa} {\overleftarrow{\bm{a}}}

\begin{document}
\title{Inference in non-equilibrium systems from incomplete information: the case of linear systems and its pitfalls}
\author{D. Lucente$^{2}$, A. Baldassarri$^{1,2}$, A. Puglisi$^{1,2}$, A. Vulpiani$^{1}$, M. Viale$^{1,2}$}
\affiliation{
	$^1$ Dipartimento di Fisica - Universit\`a La Sapienza - 00185 Rome, Italy \\
	$^2$ Istituto dei Sistemi Complessi - Consiglio Nazionale delle Ricerche, 00185 Rome, Italy \\
	}
	\date{\today}
	\begin{abstract}
	Data from experiments and theoretical arguments are the two pillars sustaining the job of modelling physical systems through inference. In order to solve the inference problem, the data should satisfy certain conditions that depend also upon the particular questions addressed in a research. Here we focus on the characterization of systems in terms of a distance from equilibrium, typically the entropy production (time-reversal asymmetry) or the violation of the Kubo fluctuation-dissipation relation. We show how general, counter-intuitive and negative for inference, is the problem of the impossibility to estimate the distance from equilibrium using a series of scalar data which have a Gaussian statistics. This impossibility occurs also when the data are correlated in time, and that is the most interesting case because it usually stems from a multi-dimensional  linear Markovian system where there are many time-scales associated to different variables and, possibly, thermal baths. Observing a single variable (or a linear combination of variables) results in a one-dimensional process which is always indistinguishable from an equilibrium one (unless a perturbation-response experiment is available). In a setting where only data analysis (and not new experiments) is allowed, we propose - as a way out - the combined use of different series of data acquired with different parameters. This  strategy works when there is a sufficient knowledge of the connection between experimental parameters and model parameters. We also briefly discuss how such results emerge, similarly, in the context of Markov chains within certain coarse-graining schemes. Our conclusion is that the distance from equilibrium is related to quite a fine knowledge of the full phase space, and therefore typically hard to approximate in real experiments.
	\end{abstract}
	\maketitle
\section{Introduction}
  
Inference from the  knowledge  of only partially accessible information is  a central challenge in  many  physical fields.
We can say that such a topic is an unavoidable link between  the experimental science and the theoretical
approach in terms of mathematical description as well as for the building of effective models from data~\cite{baldovin2018role}.
A paradigmatic example of the  importance to have  an efficient inference protocol is the problem of the phase space reconstruction in chaotic systems. 
Typically, experimental measurements provide just a time series of one observable $y$  sampled at discrete times $t_1, t_2,\dots,t_m$, so we  have a time series $y_1, y_2,\dots,y_m$,  depending on the (unknown) $D$ dimensional state vector $x=\{x_i\}_{i=1,D}$ of the underlying system.
The problem of phase space reconstruction consists in computing, from this series, quantities such as Lyapunov exponents  or  to assess 
the deterministic or stochastic nature of the system,  as well as  to build up from the time series a mathematical model enabling predictions.
Takens was able to show, using the embedding method, how, for a deterministic system under quite general assumptions,  a time series of the vector $\{y_k\}_{k=1,m}$  allows to
faithfully reconstruct the properties of the underlying dynamics~\cite{packard1980geometry,takens1981detecting,vulpiani2009chaos}. 
Such a result has a tremendous conceptual, as well as practical,  relevance: it is a  bridge between experiment and theory.
On the other hand,  even important results have their practical difficulties:
it is now well known that there are rather severe limitations for the application of the Takens method  in high dimensional systems ~\cite{ruelle1990,ferretti2020}.

Inference is rather a wide topic, and its relevance in
statistical mechanics is well recognised~\cite{baldovin2018role}.
Let us briefly introduce the problem in general terms:
we have a  system whose state is determined  by 
 a vector $x \in \mathbb{R}^D$, or an integer $i \in I_N \equiv \{1,\dots,N\}$. 
In the  first case the evolution rule is given by a differential equation (or a map), possibly including noise; for the second   case  typically one has a Markov Chain or Master equation.
In many interesting systems we have $N,D \gg 1$, in addition  some variables can be  much faster
than others; usually  in the experiments it's  impossible to  observe the whole
state of the systems but only a part.
So instead of $x \in \mathbb{R}^D$ or  $i\in I_N$
sometimes  we  can  prefer, or  are forced, to deal only with a vector $y = y(x) : x\in\mathbb{R}^D \to y\in\mathbb{R}^d $ or an integer $a = a_i : i\in I_N \to a\in I_M$ with $d<D$ and $M<N$ (and sometimes $d\ll D$ and $M \ll N$).

The  $y$s or  $a$s can be  the result of some coarse graining, projection or decimation procedure, due to the particular measurement protocol. At  a  theoretical level the natural question is how to build an effective description for $y$ or  $a$, an important topic treated in many works~\cite{pigolotti2008coarse,puglisi2010entropy,altaner2012fluctuation,bo2017multiple,teza2020exact}.
However the present paper is not devoted to such an aspect.
On  the contrary, we focus on a question which is relevant from an  experimental point of view. Namely  we  wonder whether/how from  a time series, $y$ or  $a$, it is possible to understand the original problem, or at least some of  its salient features.

Before going on, let us briefly summarize some results which, at first glance can appear negative, but have their relevance showing how, in the building of models, it is vital to adopt a pragmatic approach.
We already mentioned the crucial contribution of Takens in the analysis of the experimental data of  chaotic systems, and the severe limits in many practical cases. One of these limits is that  such a result does not hold for noisy systems~\cite{baldovin2020understanding};
the basic reason is that  noise  can be  seen as a function of time with an infinite number of Fourier terms,
and therefore it is not possible to apply to it the embedding technique using a finite dimensional vector. 
For instance when $x=(x_1,x_2,x_3) \in \mathbb{R}^3$  is a Gaussian process,  the knowledge of an even very long time series of $y=(x_1,x_2) \in \mathbb{R}^2$, in general is not  sufficient to understand the features of $x$.

Another practical limit is in the resolution of the data. An arbitrarily fine resolution of the state of the system could allow one to determine whether a given experimental signal (i.e. a time series of an observable) originates from a chaotic deterministic or stochastic dynamics, with the help of methods from information theory combined with the Takens  approach~\cite{cencini2000chaos}. 
However such a distinction is strongly limited by the practical impossibility to reach an arbitrarily fine resolution. Given that, it becomes very useful to perform an entropic analysis of a given data record in terms of $\epsilon$-entropies (and associated finite size Lyapunov exponents) which characterise the entropy of data at different resolution scales $\epsilon$~\cite{abel2000exit}. Such a result has its practical relevance: it allows us a resolution-dependent classification of the stochastic or chaotic character of a signal. In practice,
without any reference to a particular model, one can define the notion of deterministic or chaotic behavior of a system on a certain range of scales~\cite{cencini2000chaos}.

In the present paper we face a different inference  question, relevant in the context of non equilibrium statistical mechanics~\cite{peliti2021stochastic}, for which we give a brief introductory example here.
Let us assume that we know only the time series of a unique variable $y$ of a system  and we know that the underlying dynamics is a Gaussian process, e.g, $y$ is a component or a linear combination of the components of a vector $x \in \mathbb{R}^D$ whose evolution law is ruled by the linear Langevin equation
$\dot{x}  + Ax = B\xi$. Of course, a well designed perturbation-response experiment can  tell us if  the underlying system is at thermodynamic equilibrium or not~\cite{kubo2012statistical}. However such a test requires observing the system at least in two states: the unperturbed and the perturbed ones. 
A quite natural question in an experiment emerges: is the knowledge of one component, e.g. $y$,  enough to understand  that the system has a non equilibrium character~\cite{rupprecht2016fresh,seifert2019stochastic}? Note that there is, for Markov processes, a neat connection between the violation of Kubo relation (or "equilibrium" Fluctuation-Dissipation relation, EFDR) and the presence of an asymmetry under time reversal, typically measured as non-zero entropy production~\cite{harada2005equality}.
However, previous results have shown that for linear systems the above question may have negative answer~\cite{weiss1975time,diks1995reversibility,zamponi2005fluctuation}. For instance, in the case with  $x =(x_1, x_2) \in \mathbb{R}^2$  ruled by a set of linear Langevin equations with two different temperatures in such a way that the entropy production is positive,  the dynamics  of $y$ has always zero entropy production even if it does not satisfy the EFDR~\cite{crisanti2012nonequilibrium}. Such a result is rather disappointing: the  knowledge of an even very long time-series $y$ coming from the system does not  allow to understand a qualitative and fundamental aspect of the system. 
 
In this paper we put  this problem in a wider perspective, considering the class of linear systems, Markovian and non-Markovian, in full generality, and we try to suggest strategies to solve this problem in practice, from the point of view of experimental measures. The question we address has received a lot of attention in the recent years, for several categories of systems, particularly systems with a discrete space of states. For this reason we briefly review some of the strategies discussed in the literature. 

The rest of the paper is organised in the following way: in Section~\ref{examples} we consider a few explicit examples which highlight the impossibility of inferring the thermodynamic state of a system from partial observations. We apply different analysis upon the data, including conditional averages and excursion analysis: all procedure report perfect symmetry under time-reversal, when they are applied to a single component, regardless of whether the underlying two-dimensional system is at equilibrium or not. In Section~\ref{nogo} we demonstrate that this is a general limit intrinsic of linear systems, i.e. it is impossible to distinguish equilibrium from non-equilibrium looking at a time-series of a scalar observable which is a linear function of state space variables without performing response experiments. In Section~\ref{wayout} we finally show a procedure which is successful in the equilibrium/non-equilibrium distinction, but requires - in fact - the availability of data taken with different parameters, a condition which could be met in experiments even without a direct control of the parameters. The Section,~\ref{markov}, is a brief discussion of the extension of this problem to Markov chains with simple topologies (rings), motivated by the observation that very similar results apply. Conclusions and perspective are discussed in Section~\ref{conclusions}.  Appendix A contains a brief discussion of the generality of the reduction of entropy production under coarse-graining. Appendix B gives a detailed treatment of Gaussian stochastic systems, in order to make self-consistent the paper.
Finally, Appendix C shows the algorithm we use to perform numerical simulations of Ornstein-Uhlenbeck processes.
\subsection{Brief review of recent approaches}
From a statistical mechanics point of view, the lack of thermodynamic equilibrium is measured by the time-reversal asymmetry, typically measured by entopy production (EP), whose most straightforward definition has been given for Markov processes in~\cite{lebowitz1999gallavotti}. From an informational point of view,  it coincides with the Kullback-Leibler (KL) divergence between the probability of time-forward and time-backward sequences of positions in the full phase space~\cite{kawai2007dissipation,andrieux2007entropy}. In principle such a definition can be applied also to time-series of observables which bear partial information about the system, but it can only result in a lower bound to the EP, see Appendix A.

The need for a better understanding of stochastic thermodynamic from an inferential point of view has emerged in the last years~\cite{seifert2019stochastic}. Part of such an interest has been triggered by the discovery of Thermodynamic Uncertainty Relations (TURs)~\cite{barato2015thermodynamic,gingrich2016dissipation,dechant2020fluctuation,horowitz2020thermodynamic} which give, under quite general conditions, a bound to the total EP of a system based upon the knowledge of fluctuations of any partial current. This observation has led to recipes for the estimate of EP  from incomplete information, such as in~\cite{li2019quantifying,manikandan2020inferring}. It has been also shown that TUR-based approaches are usually more powerful, i.e. they give closer estimates of EP, than measuring the Kullback-Leibler information of the available partial data~\cite{roldan2021quantifying}. Given the fact that a TUR-derived bound can be improved by looking for optimal currents, machine learning approaches have also been proposed~\cite{otsubo2020estimating,kim2020learning}.  The use of TURs for the estimate of EP is certainly promising, however it requires the measurements of currents, i.e. of observables which are already indicating some breaking of time-reversal symmetry, and it is usually hard to evaluate the tightness of the obtained bound, i.e. one may obtain arbitrarily low estimates~\cite{busiello2019hyperaccurate,manikandan2020inferring}. 

We remark that currents are frequently measured by collecting a number larger than $1$ of observables, in order to see immediately the presence of cyclical trajectories. Recently several studies have put in evidence the possibility to observe cyclical trajectories in small biological systems, measuring two or more coarse-grained observables~\cite{gnesotto2018broken}, e.g. the main Fourier modes (or principal components) of some complex organism, for instance {\em C. elegans} worms~\cite{stephens2008dimensionality},  Chlamydomonas~\cite{battle2016broken}, filaments in actin-myosin networks~\cite{gladrow2016broken} and with mammalian sperms~\cite{saggiorato2017human,maggiunp2022}. In these studies the non-equilibrium character of the system is verified but a quantitative estimate of EP is  rarely considered~\cite{roldan2021quantifying}.

Strictly related to dynamical asymmetries, is the concept of avalanche shape, which has a mathematical counterpart in the concepts of bridges and excursions. In terms of stochastic processes, an avalanche or excursion corresponds to a portion of the stochastic trajectory between two successive passages through a chosen threshold. Similarly, a "bridge" is the portion of a stochastic trajectory joining a chosen starting point to a given final one without further constraints. Both these quantities have been studied for a broad class of processes, with several applications in physics~\cite{blumenthal2012excursions,barkai2014area,majumdar2015effective,baldassarri2021universal}.  
Stochastic thermodynamics represents an ideal framework where these studies could reveal their utility, for instance in comparing an excursion from an initial to a final configuration and its time reversed counterpart. Not surprisingly, tools and results from stochastic control theory and optimal transport have very recently been transferred and adopted in stochastic thermodynamics~\cite{chetrite2015nonequilibrium}. Extending these studies to non-Markovian dynamics or for incomplete information looks intriguing. In the present paper we show how bridges and excursions appear symmetric when measured on a single time-series from a Gaussian stochastic process, even if it is non-Markovian and strongly asymmetric in full phase-space, as it happens when out-of-equilibrium.
More recent papers have addressed the problem of estimating EP, or at least discriminating if it is zero or positive (that is distinguishing between equilibrium and non-equilibrium), even when the available data do not bear any signature of currents. The distribution, or some of its moments, of the residence times in certain states have been shown to be useful constraints also when observable currents are zero, leading to inferior bounds to the entropy production, but certain assumptions are required, for instance the observed states to obey semi-Markov statistics~\cite{martinez19}, or in alternative a complex optimisation problem must be solved in order to account for all possible hidden Markov state networks compatible with the constraints~\cite{skinner21,skinner21b}. The distribution of the time elapsed between certain transitions also provides a promising approach~\cite{lynn2021decomposing,harunari22,vandermeer22}. All these approaches can be applied under the validity of specific conditions and/or result in lower bounds. A lower bound is better than nothing (provided that it is not zero), but can be frequently very far from even the correct order of magnitude of EP, particularly when the investigated system  is macroscopic: for instance the Authors of~\cite{skinner21b} when analysing an experimental time-series of "residence" times for a cow to stand or lie, can only conclude that "the cows consume at least $2.4 \times 10^{-21}$ Cal/h, in
deciding whether to lie or stand". The study of times between transitions is also important in the presence of strong non-linearities, e.g. potentials with several local equilibria (multi-wells)~\cite{martin2001compressive,roldan2021quantifying,hartich2021emergent}.
Another quantity which is not directly related to time-reversal asymmetry, emerged in the study of non-equilibrium fluctuation-dissipation relations, is dynamical activity (or sometimes "frenesy"), which is also known to provide bounds to entropy production~\cite{maes17,terlizzi18}.
A different interesting strategy is to compare data coming from regimes realised with different choices of a certain parameter~\cite{polettini2017effective}. This approach is, for some aspects, similar to a response experiment, but one could imagine that such parameter-dependent sets of data are already available and can be exploited in order to realize a kind of “a posteriori '' (in principle non-linear) response experiment. For instance, this situation could be realized even if the observer cannot directly influence the parameters of the system (e.g. in weather or climate observations). We will apply similar ideas to our problem in Section~\ref{wayout}.

\section{Pitfalls of linear systems}
\label{examples}
Let's first introduce a very simple example of stochastic linear system, which in its most general form is called Brownian Gyrator, a model recently adopted to describe experimental systems and nano-machines~\cite{ciliberto2013heat,ciliberto2013statistical,argun2017experimental,cerasoli2022spectral}. It consists of a linear two-variables Markovian system whose evolution is ruled by the following stochastic differential equation~\cite{filliger2007brownian}
\begin{equation}\label{eq:BG}
\left\{
\begin{array}{l}
\dot{x}_1 = -\alpha x_1 + \lambda x_2 + \xi_1 \\
\dot{x}_2 = -\gamma x_2 + \mu x_1 + \xi_2 \\
\end{array}
\right.
\end{equation}
where the $\xi$'s are two independent noise sources, whose amplitudes may differ $\ave{\xi_i(t)\xi_j(t')}=2T_i\delta_{ij}\delta(t-t')$. It is important to remark that many of our examples are good descriptions of overdamped Brownian particles in contact with baths and therefore it is reasonable (assuming damping coefficients to be $1$) to identify the noise amplitudes $T_1$ and $T_2$ as temperatures. (Here we are interested in the case where $\alpha>0, \gamma>0, \alpha\gamma>\lambda\mu$, where the system asymtptotically converge to a stationary distribution).

\begin{figure*}
\centerline{\includegraphics[scale=0.35]{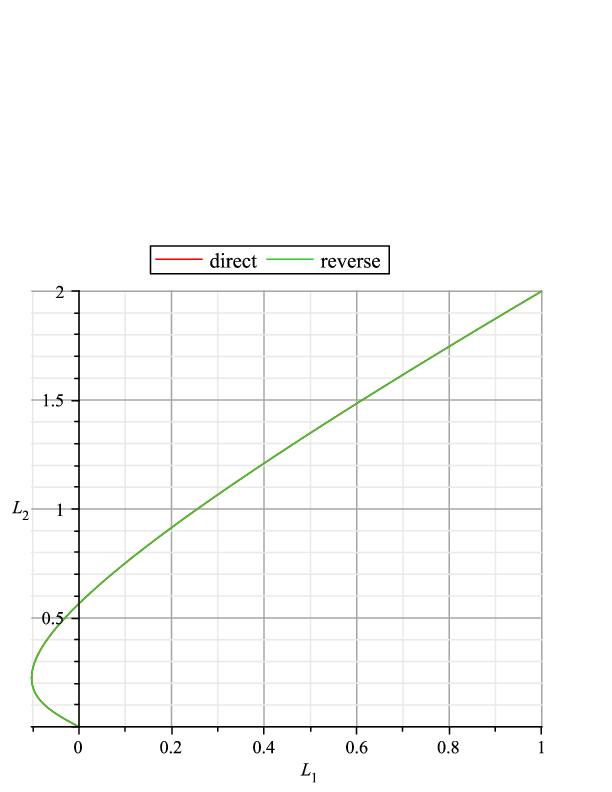}\includegraphics[scale=0.315]{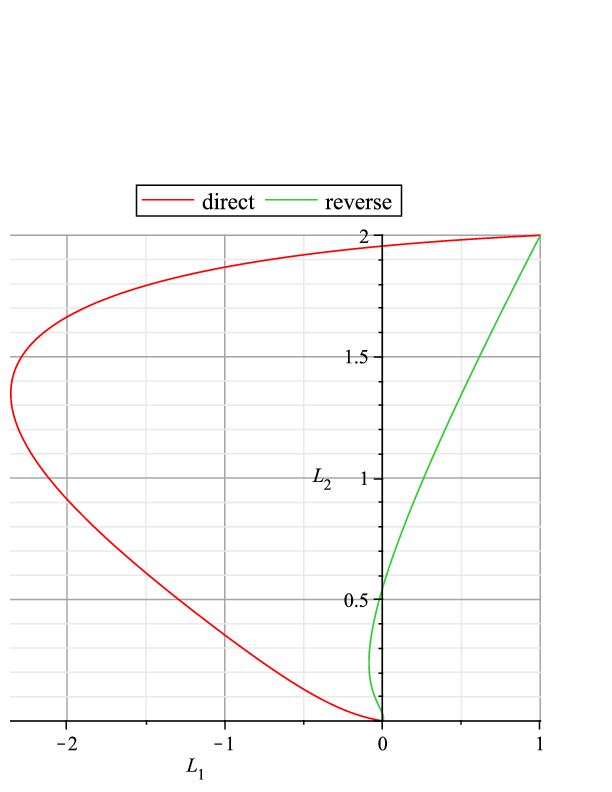}}
\caption{Average trajectory for Brownian Gyrator (drift paramters $\alpha=\gamma=1$, $\lambda=\mu=-0.5$) for trajectories of duration $\tau=5$, from $x_i=(0,0)$ to $x_f=(1,2)$ and viceversa. The two panels represent different choices of temperatures $T_1, T_2$: at left the system satisfies detailed balance ($T_1=T_2=2$), at right detailed balance is broken ($T_1=10, T_2=1$). These are parametric plots, where the axis are the two average components of the bridge $L_i(t) = \langle x_i(t)\rangle^{x_i\to x_f}_\tau$, while time $t$ is the parameter.}
\label{fig:BridgeTrajectory}
\end{figure*}

In order to illustrate the behaviour of the model, let's consider the average trajectory between two points in the configuration space. In Fig.~\ref{fig:BridgeTrajectory}, it is shown the average trajectory between the point $x_1=0, x_2=0$ and $x_1=1, x_2=2$, for trajectory of fixed duration $\tau$. In each panel we show the average trajectory for the direct $(0,0)\to(1,2)$ and the reverse $(1,2)\to(0,0)$ path. Not surprisingly, for equal temperatures $T_1=T_2$ (left panel) the direct and the reverse path coincide. This is a consequence of detailed balance, which holds for an equilibrium system. On the other hand, when $T_1 \neq T_2$ (right panel) detailed balance is broken, and this has an immediate consequence on the average direct and reverse trajectories, which now take two completely different paths. The figure gives a very intuitive visualization of the presence of the internal probability current, characterizing the non-equilibrium stationary state for the general $T_1\neq T_2$ case.

The problem we are interested in is wether is possible to appreciate the non equilibrium nature of the process, having access to the information given by one component only (say $x_1$). Given the simplicity of the model, is it possible to carry out several exact computations. In fact the quantities shown in Fig.~\ref{fig:BridgeTrajectory} are related to the the so-called "bridge" of the stochastic process.

The bridge is the process obtained constraining the trajectories of the original stochastic process to some initial and final (after a chosen time $\tau$) configurations. In other words, the bridge is the ensemble of trajectories which connect $x_i$ to $x_f$ in a time $\tau$. 
The probability to observe a value $x$ of the bridge at an intermediate time $0<t<\tau$ is given by
\begin{equation}
\cP_b\left(x,t\, |\, x_i, x_f ,\tau\right) = \frac{\cP(x,t\,|\,x_i,0) \cP(x_f,\tau-t\,|\,x,0)}{\cP(x_f,\tau\,|\,x_i,0)},\label{Bridgedefinition}
\end{equation}
where $\cP(x,t_1|y,t_2)$ is the propagator of the Markov process (which has been also considered homogeneous in time $\cP(x,t_1|y,t_2)=\cP(x,t_1-t_2|y,0)$). 

We consider this quantity as a tentative proxy of the equilibrium status of the process, since it can be shown (for variables which are even under time-reversal) that detailed balance implies the following symmetry of the bridge distribution:

\begin{equation}
\cP_b(x, t \,|\,  x_i, x_f ,\tau) = \cP_b(x, \tau-t \,|\,  x_f, x_i ,\tau). \label{BridgeSymmetry}
\end{equation}

In the case of a Brownian Gyrator, because of the linearity of the process, the bridge is a Gaussian (non homogeneous) process, and the distribution Eq.~(\ref{Bridgedefinition}) is a multivariate Gaussian distribution, fully characterised by its mean vector $L(t)$ and its covariance matrix $Q(t)$. Their expressions are~\cite{Chen2016}:
\begin{eqnarray*}
L(t) &=& \cR(t) x_i + P(t) \cR(\tau-t)^T P^{-1}(t) \left[ x_f - \cR(\tau)x_i\right] \\
Q(t) &=& P(t) - P(t) \cR(\tau-t)^T P^{-1}(\tau)\cR(\tau-t)P(t),
\end{eqnarray*}
where $\cR(t) = \exp(-A t)$ is the solution of the deterministic equation $\frac{d \cR(t)}{dt} = -A \cR(t)$ with initial condition $\cR(0) = I$ (i.e. the response, see Appendix B), and $P(t)$ is the covariance of the propagator, which satisfy the Lyapunov equation $\frac{dP}{dt} = -A P - P A^T + \Sigma$ with initial condition $P(0)=0$ and where $\Sigma_{ij} = 2 T_i \delta_{ij}$. (The covariance of the propagator can be also expressed in terms of correlation $\cC$ and response $R$ of the process as $P(t) = \cC(0) - R(t) \cC(0) R(t)^T$).

As before mentioned, Fig.~\ref{fig:BridgeTrajectory} shows the mean $L(t)$ (the plane coordinates corresponds to the coordinates of the vector $L_1(t)$ and $L_2(t)$) for a bridge of a Brownian Gyrator from $x_i=(0,0)$ to $x_f=(1,2)$ compared with the mean of the bridge with endpoints reverted $x_i=(1,2)$ and $x_f=(0,0)$. As can be seen, because of Eq.~\ref{BridgeSymmetry}, when detailed balance is satisfied ($T_1=T_2$) the two paths (direct and reverse) coincide, while they differ for $T_1\neq T_2$.

In the present work, we are not interested in the information contained in the full bridge distribution Eq.~(\ref{Bridgedefinition}). Rather we consider the case where only a single component of the Brownian Gyrator is accessible by measures. A single component is still a Gaussian, stationary process, but it is no more Markovian (since the other variable is now hidden). 

Nevertheless, one can consider the moments of the component of the bridge:
\[
\langle x_1(t)^n\rangle^{x_i\to x_f}_\tau = \int dx\, x_1^n\, \cP_b(x,t\,|\,x_i,x_f,\tau),
\]
which, for $n=1$, recovers the first component of the mean vector $L(t)$, while for $n=2$ is $P_{11}(t)+L_1(t)^2$. Note that the moments depend on $\tau$ and on the extreme points $x_0$ and $x_f$, which include both components of the original Markovian process.

\begin{figure*}
\centerline{\includegraphics[scale=0.35]{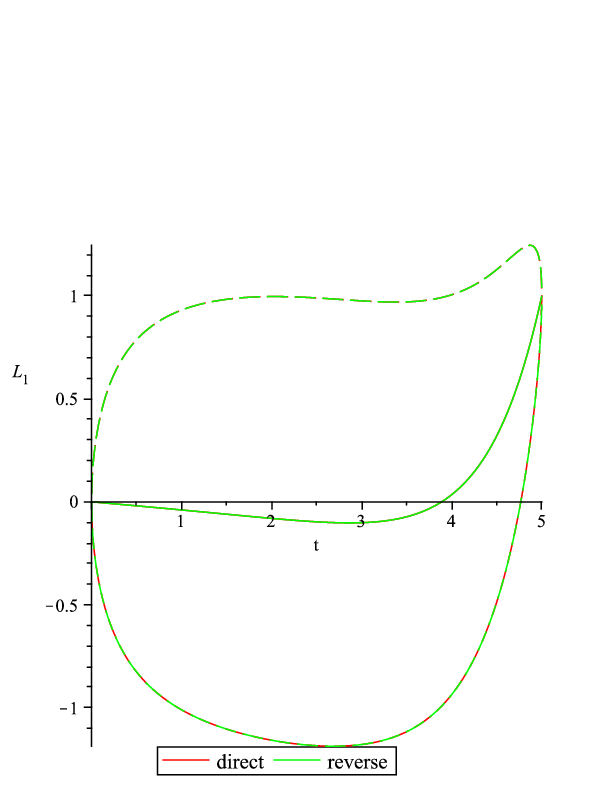}\includegraphics[scale=0.35]{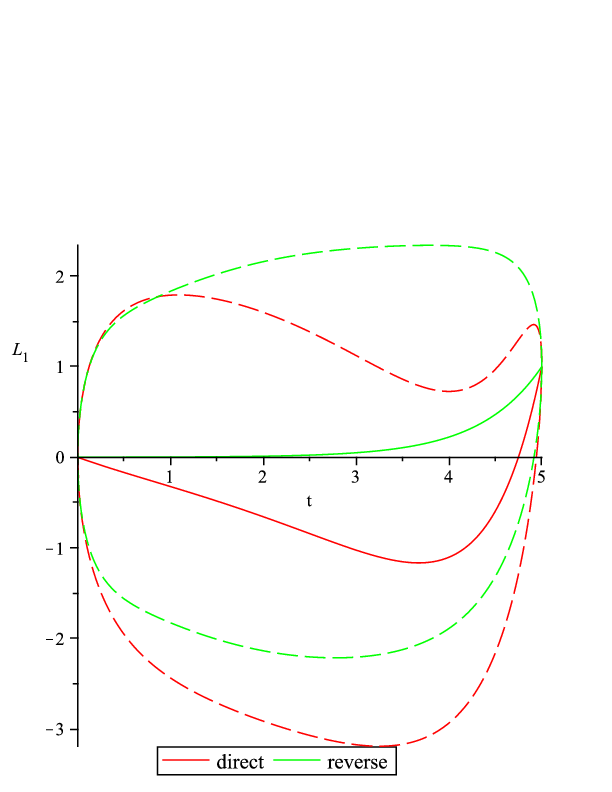}}
\caption{Single component of the average bridge trajectories of Fig.~\ref{fig:BridgeTrajectory}, as a function of $t$. The dashed lines correspond to $\pm \sqrt{Q_{11}}$, and give a better ideas of the fluctuations of the bridge trajectories. Same parameters of Fig.~\ref{fig:BridgeTrajectory}. Again, left panel is a case with detailed balance ($T_1=T_2$), right panel is the out-of equilibrium case ($T_1\neq T_2$.}
\label{fig:BridgeFirstComponent}
\end{figure*}

Again, if detailed balance is satisfied, the simmetry Eq.~\ref{BridgeSymmetry} implies:
\begin{equation}
\langle x_1(t)^n\rangle^{x_i\to x_f}_\tau = \langle x_1(\tau-t)^n\rangle^{x_f\to x_i}_\tau, \label{ShapeSymmetry}
\end{equation}
that means that the moments have symmetric shape with respect to $t=T/2$.
In Fig.~\ref{fig:BridgeFirstComponent}, we show $L_1(t)$, as well as $L_1(t)\pm \sqrt{Q_{11}(t)}$ for the same bridges considered in Fig.~\ref{fig:BridgeTrajectory}. Again, in the case of detailed balance ($T_1=T_2$), moments satisfy the time reverse symmetry Eq.(\ref{ShapeSymmetry}). However, in the non equilibrium case ($T_1\neq T_2$), the symmetry is broken.

\begin{figure*}
\centerline{\includegraphics[scale=0.35]{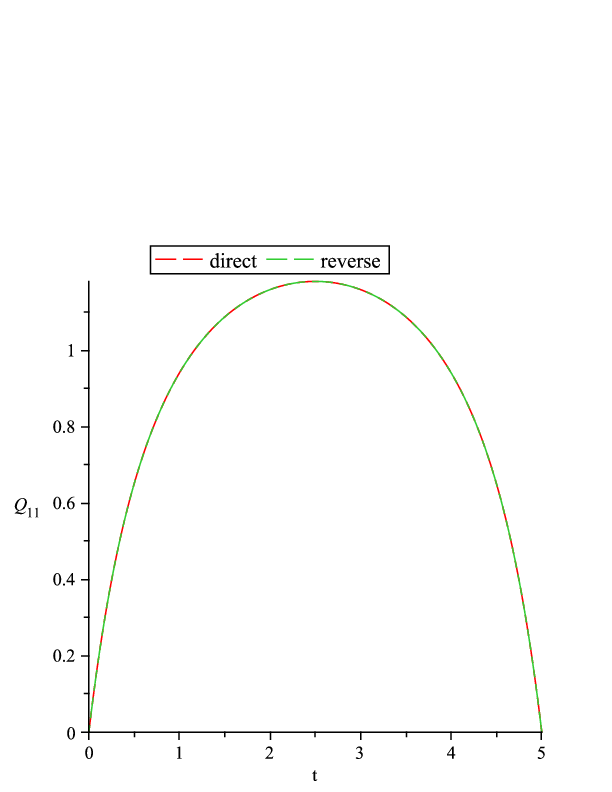}\includegraphics[scale=0.35]{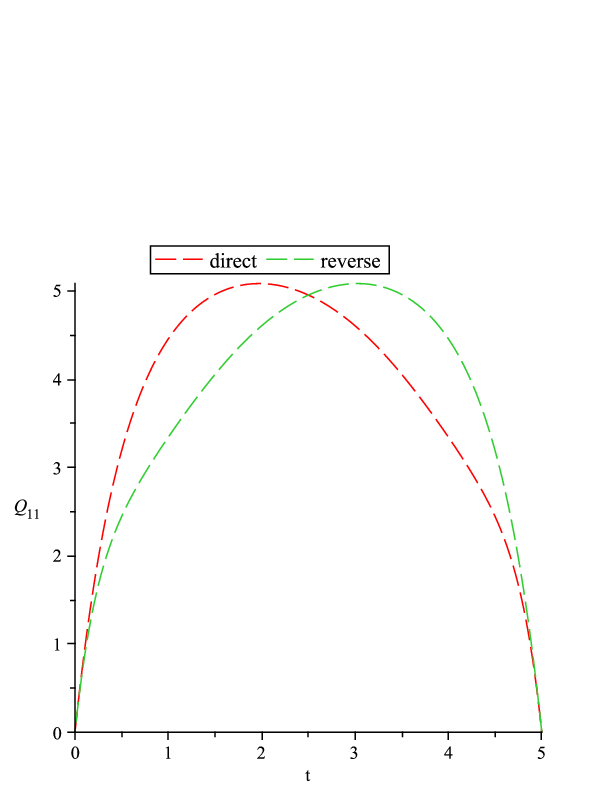}}
\caption{In the case of a bridge from $x_i=(0,0)$ to $x_f=(0,0)$ after a time $\tau=5$, the values of the covariance matrix element $Q_{11}$ are shown as a function of time. For the equilibrium case ($T_1=T_2$, left) the curve is symmetric around $t=\tau/2$, and the direct curve corresponds exactly to the reverse bridge. On the contrary, out of equilibrium ($T_1\neq T_2$), direct and reverse path differ, and the curves are no more symmetric. The values of the paramteres are the same of Fig.~\ref{fig:BridgeTrajectory}}
\label{fig:BridgeSigmaXX}
\end{figure*}

In particular, one can consider the special bridges with $x_i=x_f=(0,0)$: in this case, since one has obviously $L(t)=0$, one can measure the second moment, which correspond to $Q_{11}(t)$. In Fig.~\ref{fig:BridgeSigmaXX} we show the behaviour of $Q_{11}(t)$ for such a bridge, with and without detailed balance.

Interestingly, figures as Fig.~\ref{fig:BridgeSigmaXX} are very similar to the average avalanche shape, a measure introduced in the context of the study of crackling signals~\cite{Sethna2001,Mehta2002}, in the context of Barkhausen noise in ferromagnetic materials~\cite{Alessandro1990,Alessandro1990b}. In this case, the signal analized is a (positive) bursty measure. Given a small (ideally zero) threshold, the signal is regarded as a sequence of avalanches (the signal between to successive zeroes). In practice the avalanche represents a single burst of activity of the signal, between to quiescent phases (think for instance to the intensity of a single earthquake). Then, avalanches of the same durations are averaged in order to get the average avalanche shape.

In terms of the theory of stochastic processes, the avalanche of the process is called an excursion~\cite{Pitman2018,majumdar2015effective}. In a recent paper~\cite{baldassarri2021universal}, for the case of a class of multiplicative stochastic processes (ABBM/CIR/Bessel processes), it has been shown that the average bridge shape is simply proportional to the average avalanche shape, suggesting that the two quantities (bridge and excursion) carry similar information about the time evolution of the process. Symmetric, as well asymmetric average avalanche shapes has been observed in several physical~\cite{Bares2014,Bares2017,Baldassarri2019,Vu2020}, geophysical~\cite{Mehta2006} and biological~\cite{Miller2019} phenomena, but at the moment there is no general understanding of the meaning of such property. The only work~\cite{Durin2007} devoted to the asymmetry of some Barkhausen average avalanche shape attributes the phenomenon to subtle inertial effects of the effective motion of magnetic domains inside the ferromagnetic material. 

Coming back to the bridge shapes of the Brownian Gyrator, Fig.~\ref{fig:BridgeSigmaXX} could give the illusion to represent a measure on a single component, which could discriminate the equilibrium nature of the whole process: the equilibrium case corresponding to a symmetric shape, while the off-equilibrium an asymmetric one.

Unfortunately, the quantity considered includes more information with respect to the single component, since it represent the average shape of a bridge comprised by two points where both the coordinates of the full process are zero. In other terms, in order to perform such a measure, one needs an information on both coordinates.

In order to consider the more general case, where one has access strictly to a single component, one have to consider the stationary bridge of such a component, independently from the value of the second component.
This turns to a different definition of the bridge, with respect to Eq.~(\ref{Bridgedefinition}).
In fact, once fixed the values of the first components of $x_0$ and $x_f$, one has to perform a stationary average over the second component of $x_0$ and then integrate over every possible values of the second component of $x_f$ and $x$. More precisely, the bridge distribution for the stationary first component, going from $x_1(0)=x_{1i}$ to $x_2(\tau)=x_{2f}$ is:
\begin{widetext}
\begin{equation}
\cP^{(1)}_{b}(x_1,t|x_{1i},x_{1f},\tau) = \frac{\int \,dx_{2i} \,dx_{2f}\, dx_2 \cP^s(x_i)\cP(x,t\,|\,x_i,0)P(x_f,\tau\,|\,x,t)}{\int\, dx_{2i}\, dx_{2f} \cP^s(x_i) \cP(x_f,\tau \,| \,x_i,0) },\label{BridgeSingle}
\end{equation}
\end{widetext}
where 
\[
\cP^s (x) = \lim_{t\to \infty} \cP(x,t | y,0),
\]
is the stationary distribution of the (free) process.

\begin{figure*}
\centerline{\includegraphics[scale=0.35]{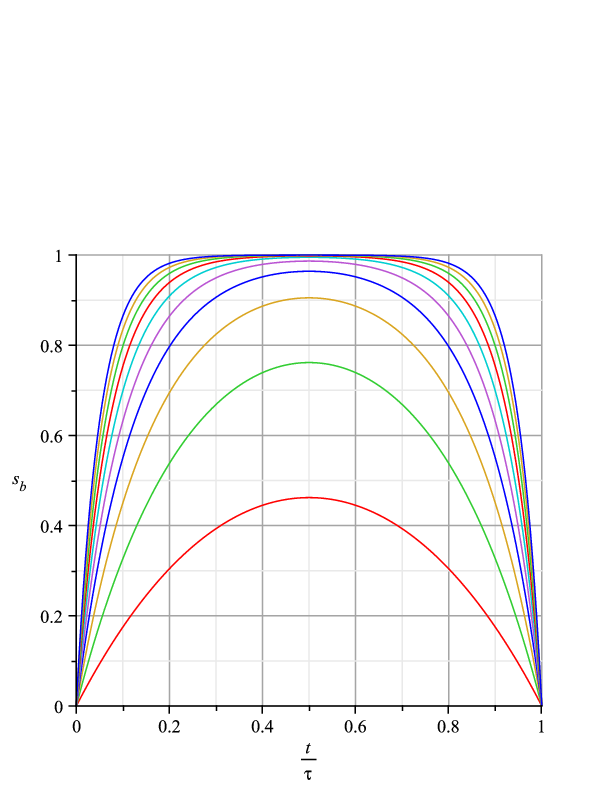}\includegraphics[scale=0.35]{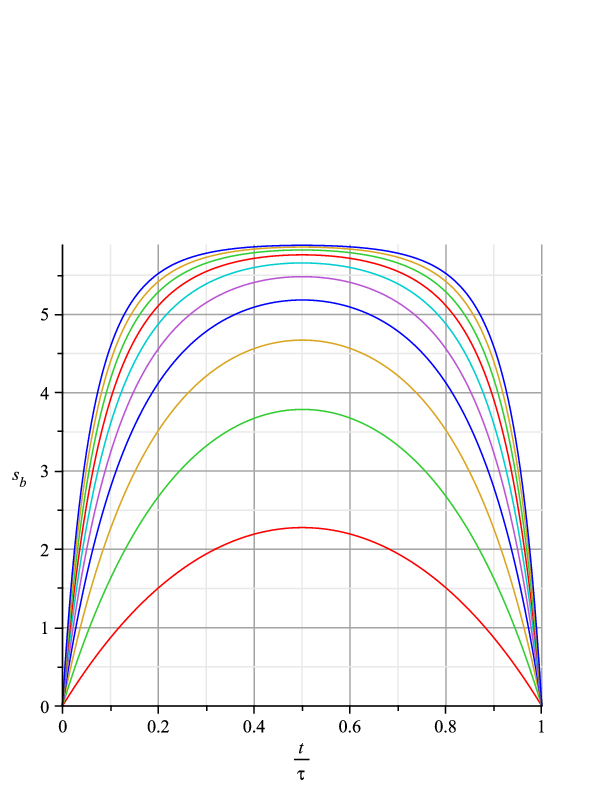}}
\caption{Average bridge shape for a stationary single component of a Brownian Gyrator, with (left) and without (right) detailed balance. In each panel, the curves represent different duration values $\tau=1,2,..,10$, from bottom to top. The parameters of the Gyrator are the same of Fig.~\ref{fig:BridgeTrajectory}.}
\label{fig:CompareBridgeSingleComponent}
\end{figure*}

Due to gaussianity of the process, the computation of~(\ref{BridgeSingle}) for the Brownian Gyrator can actually be performed, but it's too cumbersome to be shown here, even in the case of the symmetric gyrator $\alpha=\gamma$ and $\lambda=\mu$. However, an example of single component bridge is shown in Fig.~\ref{fig:CompareBridgeSingleComponent}.

There we compare the average shape (its variance) of the bridge for the single component between two zero values ($x_{1i}=x_{1f}=0$), for several values of duration $\tau$:

\[
s_b = \langle x_1(t)^2 \rangle^{x_{1i}\to x_{1f}}_\tau = \int \,dx_1\, x_1^2\, \cP^{(1)}_{b}(x_1,t|x_{1i},x_{1f},\tau).
\]
 In the left panel the component comes from a Brownian Gyrator at equilibrium ($T_1=T_2$), while the right panel the system is out-of-equilibrium ($T_1\neq T_2$). In both cases, the shape is symmetric with respect to $t=\tau/2$, and there is no way to appreciate the off-equilibrium origin of the second case.

This shows that, for linear systems, the bridge of a single component can not be a proxy for the determination of the equilibrium nature of the full system. In fact, this is due to a very general mathematical results~\cite{weiss1975time}: any scalar gaussian process, not necessarily Markovian, which is statistically invariant under time translation is also statistically invariant under time reversal. 

In order to grasp a more physical intuition of such a quite suprising result, we consider more standard quantities.

Suppose we can carry out an experiment in which the only measurable quantity is the observable $y=x_1$. Since the system is linear, the variable $y$ is Gaussian and therefore its correlation function $\cC_y(t)$ completely characterizes observed stochastic process. 
Nevertheless, there is an infinity of underlying linear, markovian bidimensional systems which are consistent with the observations.

For instance, consider two systems\footnote{See Eq.\ref{eq:equivalent_process} in Sec.\ref{nogo} for a detailed discussion on how the parameters of $S_1$ have been chosen.}:

\begin{itemize}
\item{$S_1$:} defined by the parameters $\alpha\simeq2.466$, $\gamma\simeq0.8667$, $\lambda=\mu\simeq-0.8969$, $T_1=T_2=\frac 12$ 
\item{$S_2$:} defined by the parameters $\alpha=\frac 83$, $\gamma=\frac 23$, $\lambda=\mu=-\frac 23$, $T_1=\frac 15$, $T_2=\frac 12$
\end{itemize}
%{\bf Bisogna mettere i valori giusti per la matrice $A$ del sistema $1$}

\begin{figure*}
\centerline{\includegraphics[scale=0.5]{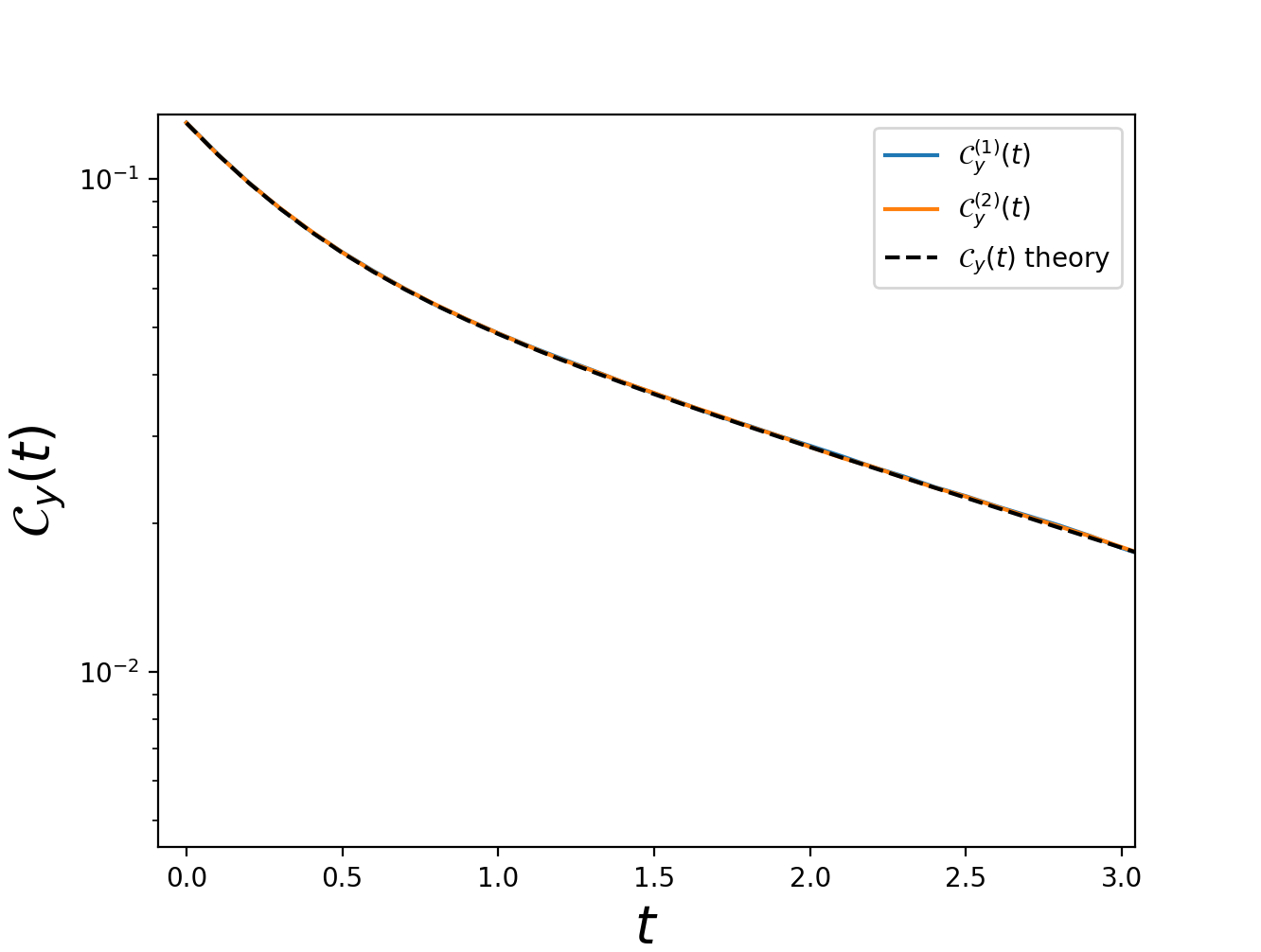}\includegraphics[scale=0.5]{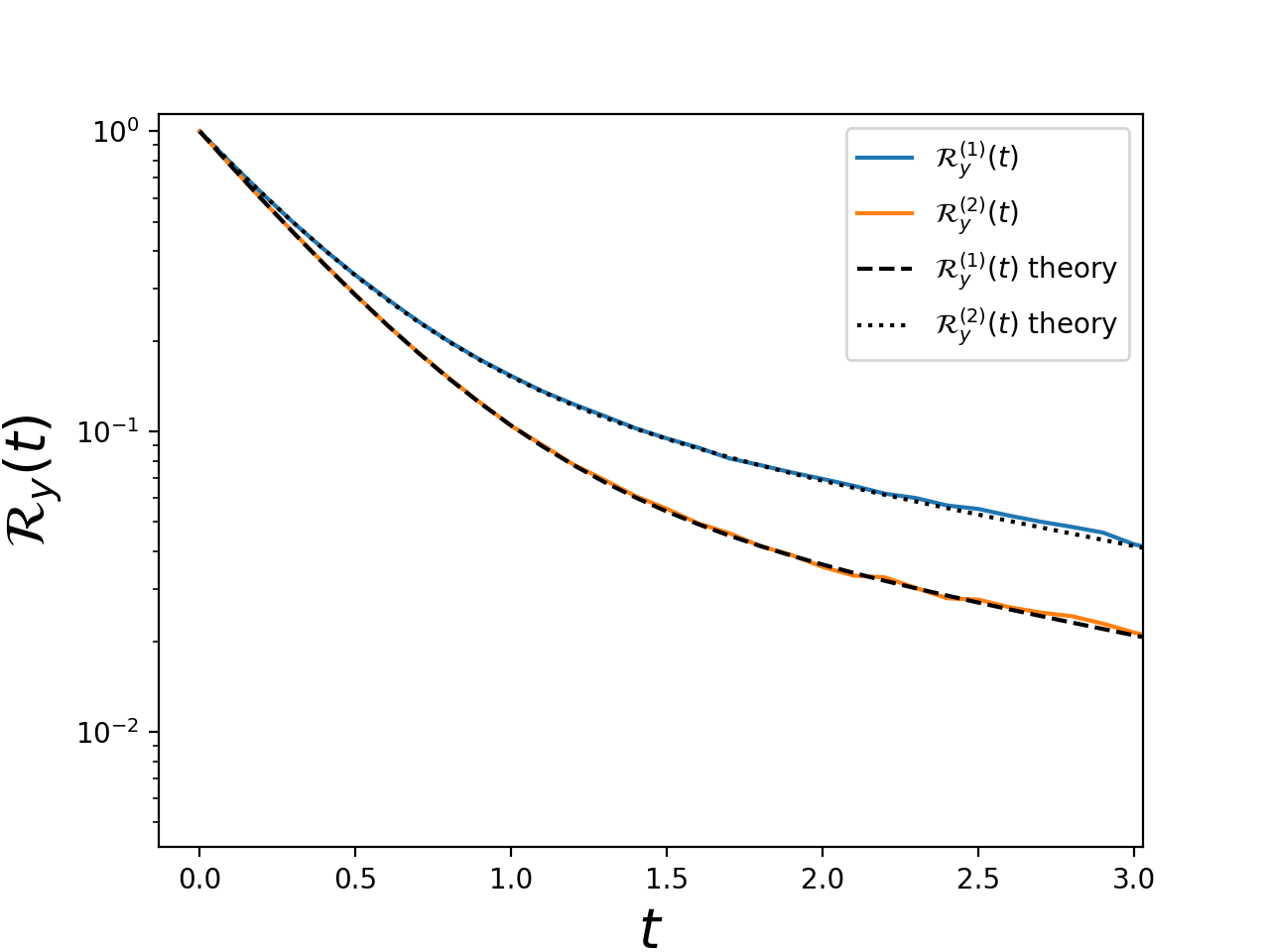}}
\caption{Left panel: autocorrelation function $\cC_y(t)$ for the two systems $S_1$ (blue) and $S_2$ (orange).  Right panel: response functions $\cR_y(t)$ of $S_1$ (blue) and $S_2$ (orange). The results have been obtained by means of numerical simulations. The analytical results are shown in black.}
\label{fig:Correlation_Response}
\end{figure*}

Note that, while $S_1$ is at equilibrium ($T_1=T_2$), system $S_2$ is not: the average value of its entropy production rate is $S=0.12$.
However, looking at $y=x_1$ only, it can be shown that both systems share exactly the same correlation function $\cC_y(t)$ 
\begin{equation}
\cC_y^{(1)}(t)=\cC_y^{(2)}(t)=\cC_y(t)=c_+e^{-l_+|t|} +c_-e^{-l_-|t|}
\end{equation}
but different response functions
\begin{align}
  \cR_{y}^{(1)}(t)&=r^{(1)}_+e^{-l_+t} +r^{(1)}_-e^{-l_-t}\qquad \qquad (t>0)\nonumber\\
  \cR_{y}^{(2)}(t)&=r^{(2)}_+e^{-l_+t} +r^{(2)}_-e^{-l_-t}
\end{align}
as shown in Fig.\ref{fig:Correlation_Response}. 

\begin{figure*}
\centerline{\includegraphics[scale=0.5]{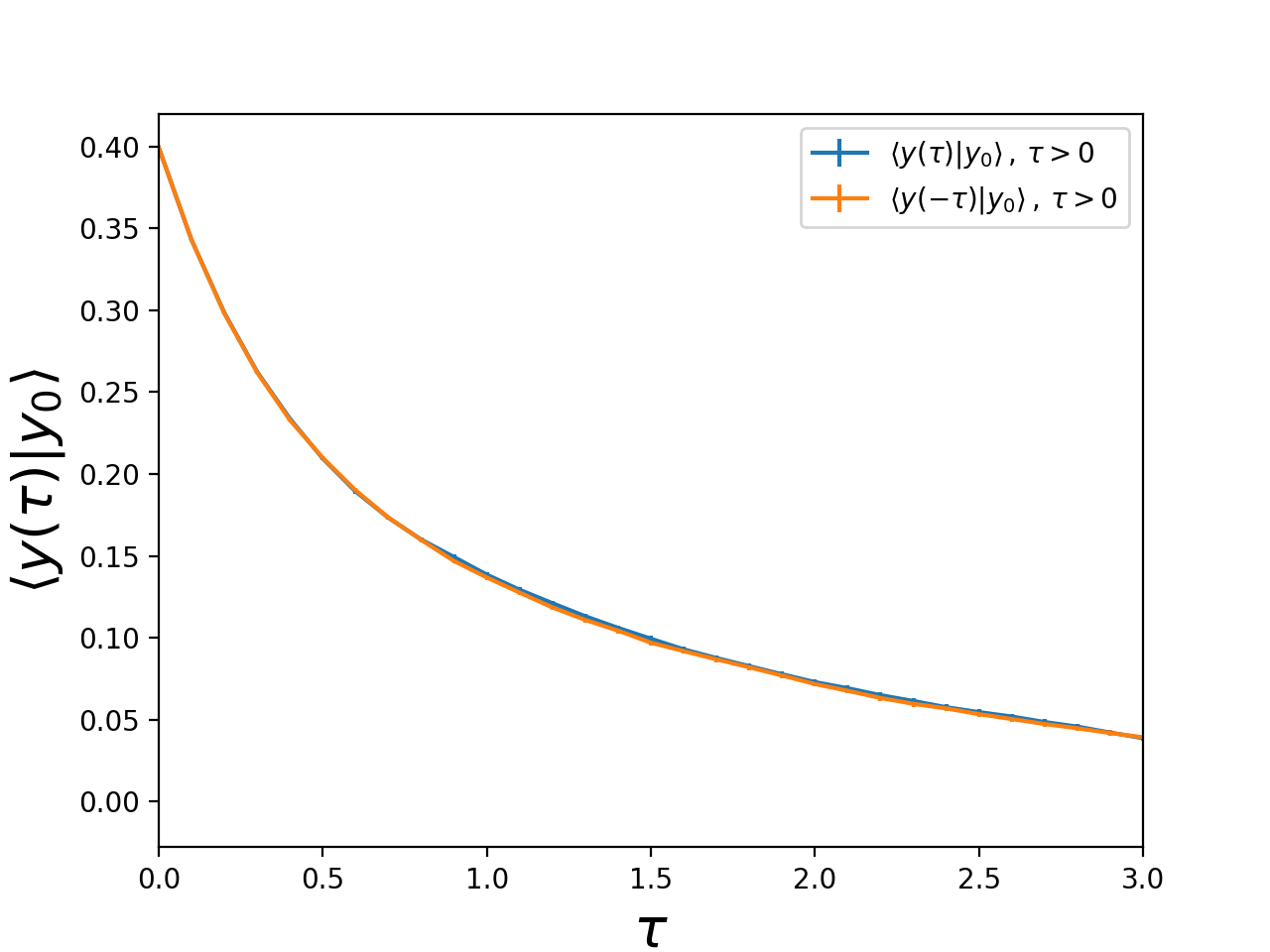}\includegraphics[scale=0.5]{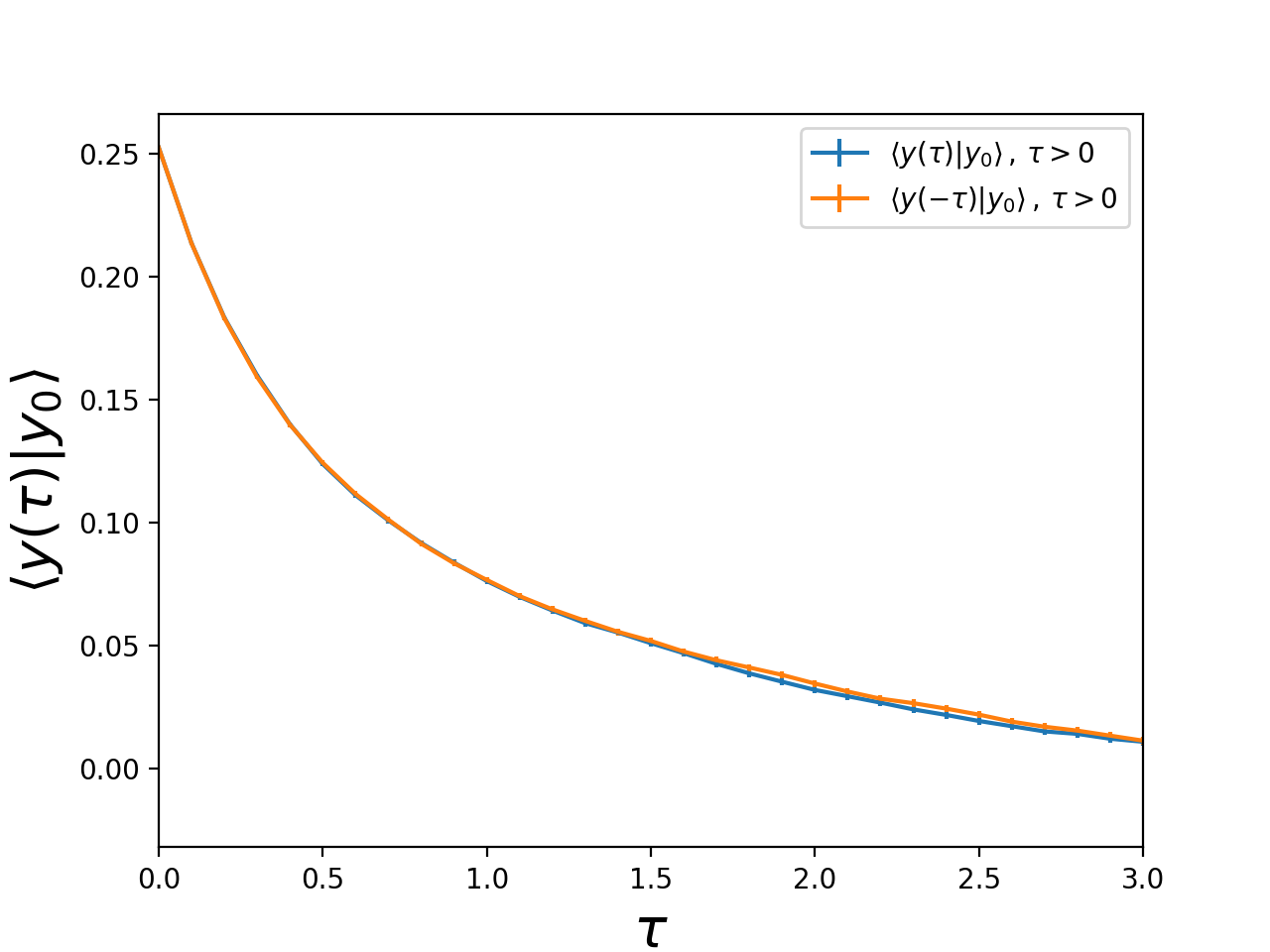}}
\caption{Conditional average $\langle y(\tau)|y_0\rangle$ as a function of the time delay $\tau$ for the two systems $S_1$ (left) and $S_2$ (right). In both cases $y_0=0.7\sigma_y$, where $\sigma_y$ is the standard deviation. The results have been obtained by means of numerical simulations.}
\label{fig:Conditional_average}
\end{figure*}

It is therefore evident that without knowing the response functions the two systems cannot be distinguished.
One might be tempted to find out whether the detailed balance is satisfied by looking at suitable statistical features. Actually, if the system is invariant under the transformation $t\rightarrow-t$ then it follows that every statistical quantity is an even function of time. The panels of Fig.\ref{fig:Conditional_average} show the conditional averages $\langle y(\tau)|y_0\rangle$ and $\langle y(-\tau)|y_0\rangle$ for the two systems. As can be seen, these objects are invariant under $t\rightarrow-t$. 
One might naively think that, despite $\langle y(\tau)|y_0\rangle=\langle y(-\tau)|y_0\rangle$, possible dissipative effects occur in statistical objects that depend on two or more times due to the non-Markovianity of the process. For instance, given $\Delta t>0$, one can consider the two quantities $\langle y(\tau)|y_0,\pm\rangle=\langle y(\tau)|y(0), y(-\Delta t)\lessgtr y(0)\rangle$. 
$\langle y(\tau)|y_0,+\rangle$ and $\langle y(\tau)|y_0,-\rangle$ are the mean values of $y(\tau)$ knowing that $y(0)=y_0$ has been reached from below or from above, respectively. In some sense, the second conditioning is equivalent to providing information also on the "velocity" of the process. For a process which is invariant under the transformation $t\rightarrow-t$ we have $\langle y(\tau)|y_0,+\rangle=\langle y(\tau)|y_0,-\rangle$ since there are no differences to look at the process forward or backward in time. 
Not surprisingly, $\langle y(\tau)|y_0,+\rangle$ and $\langle y(\tau)|y_0,-\rangle$ are indistinguishable (not shown). Hence, it is not possible to infer the thermodynamic state of the system.

In the following sections, we will provide a general proof of such indistinguishibility, based on an analysis of correlation and response functions of the system. Furthermore, we propose a possible way out, which could be used when a limited control on the system is available.

%%%%%%%%%%%%%%%%%%%%%%%%%%%%%%%%%%%%%%%  
\section{A "no-go" theorem for linear systems}

\label{nogo}

First, we assume to know - from previous knowledge - that the data are generated by a stochastic linear system (being it Markovian or not). Then, we know that the only statistical information that can be retrieved from a time series of a set of $D$ observables measure during its evolution is the stationary correlation matrix $\cC(t-t')$ whose elements are $\ave{x_i(t)x_j(t')}$.

If we are only interested in discriminating equilibrium from non equilibrium, and we are lucky enough to have access - with our available observables - to a subset large enough of the full phase space,
we can simply check if the detailed balance
is satisfied or not, by looking at the condition $S\cC(t) S=\cC(t)^T$, where $S_{ij}=s_i\delta_{ij}\; s_i\in\{-1,+1\}$ takes into account the effect of the time reversal on the different components (see Appendix B).
But what happens when the space of the real phases of the system under observation is very large and, instead, our observables are a few or even only one?

In one dimension in particular, the detailed balance condition is inevitably always satisfied, because $C(t)$ is a scalar function. If the system is non-Markovian, however, it is possible that non-equilibrium information is contained in the comparison between the time-correlation of the noise and the memory kernel function representing the deterministic force of the system, i.e. by evaluating the so-called $2$-nd kind Fluctuation Dissipation Relation~\cite{kubo2012statistical}. In order to do so, however, we need to separate the contribution of the noise from that of the deterministic forces. 

In Appendix B we show in details that such separation is impossible; here we summarise the situation.
%that with the knowledge of $\cC(t)$, we are unable to discriminate the contribution of noise from that of the memory kernel and therefore we cannot distinguish between equilibrium and non-equilibrium. 
The dynamics of our small set of observables $x=\{x_1,\dots,x_D\}$ is described by a linear integral-differential stochastic equation of the type $\cL x = \cB\xi$, where $\cL$ and $\cB$ are operator acting on a vector space of functions in $\text{L}^1(\mathbb{R})$ and $\xi$ is a (in general non-diagonal) colored noise matrix. In this case the Fourier transform of the connected correlation function $\cC_{ij}(t-t')=\ave{x_i(t)x_j(t')}_c$ is related to the linear response of the system $\lx.\ave{\partial x_i(t)/\partial h_j(t')}\rx|_{h=0}=\cR_{ij}(t-t')$ and to the noise fluctuations $\ave{\xi_\alpha(t)\xi_\beta(t')}_c = \nu_{\alpha\beta}(t-t')$ according to (see Appendix B)
\begin{equation}\label{com}
    \cCfo = 2\pi \cRfo \cSgfo \cRfo^\dagger = \cCfo^\dagger
\end{equation}
where $\sqrt{2\pi}\cRfo=\cLfo^{-1}$ and $$\cSgfo = \cBfo \widetilde{\nu}(\om) \cBfo^\dagger = \cSgfo^\dagger.$$
The equation (\ref{com}) states that in general, in the absence of any other information (in particular, without response experiments), it is impossible to separate the contribution of the noise from that of the response by simply looking at the correlation function: the noise and response poles are mixed together and we cannot assign them to one source or the other without ambiguity.
In order to show such an impossibility we can consider the following one dimensional stochastic process with a colored noise ( $\ave{\xi(t)}=0$ ) 
\bea
\dot{x} + ax = \xi \qquad \ave{\xi(t)\xi(t')}=2T e^{-b|t-t'|}
\eea
with $a>0$ and $b>0$, for which, a direct computation leads to the following expression for the Fourier transform of the time correlation function $\cCfo$ and for the response $\cR(t)$ of the system
\bea
\cCfo = \frac{2T}{(\om^2+a^2)(\om^2+b^2)} \quad \cR(t) =\theta(t) e^{-at}
\eea
Without a response experiment we are not able to distinguish the stochastic process above from the one having inverted rates 
\bea
\dot{x} + bx &=& \xi \qquad
\ave{\xi(t)\xi(t')}=2T e^{-a|t-t'|} \\
\cR(t)&=&\theta(t)e^{-bt}
\eea
or, not necessarily worse, we might think to have $\delta$-correlated white noise and a second order SDE: in this case we could have something like
\bea
\ddot{x} +(a+b)\dot{x}+(ab) x &=&\xi \qquad
\ave{\xi(t)\xi(t')}=2T\delta(t-t') \\
\cR(t) &=& \theta(t) \frac{e^{-bt}-e^{-at}}{a-b}.
\eea
Then, if we are unable to perturb the system and measure its response, then we are unable to understand what process we are really looking at.
We can hope to deal with this problem by restricting the set of systems under investigation. For instance, we can think to have a multidimensional model for variables $\{x_i\}$ ($i=1...D$) and that an equation like $\cL y=\cB \xi$ arises once we just observe a single dynamic variable $y=x_i$ or a linear combination $y=\sum_i a_i x_i$ of these or their derivatives.
A very general and natural model is a genuine multidimensional Ornstein–Uhlenbeck process as $\dot{x}+Ax=B\xi$ or its subset $\ddot{x}+\Gamma\dot{x}+Kx=B\xi$ where the dynamical variables $x$ and $\dot{x}$ have opposite time reversal parity and $\ave{\xi_i(t)\xi_j(t')} = \nu_{ij}\delta(t-t')$.
In this way, with a good data fit like
\bea
\cC_y(t) &=& \sum_\alpha c_\alpha^{(0)} \; e^{-\gamma_\alpha t} + \\ &+&\sum_\beta e^{-\gamma_\beta t} \lx(c_\beta^{(+)} \cos{\Omega_\beta t} + c_\beta^{(-)}\sin{\Omega_\beta t} \rx)
\eea
and a robust statistical analysis able to determine the number of exponentials to consider, we can hope to bet on the dimension of the hypothetical underlying multidimensional system and to recover the poles of the response function (since we assume the noise does not have it). 
In many cases this approach seems to work reasonably well when, for example, we want to distinguish a second order dynamic from a first order one.
Let's imagine that we have obtained the time correlation function $\cC_y(t)=c_+ e^{-l_+|t|} + c_- e^{-l_-|t|}$ from the evolution of an one dimensional observable $y=x_1$ which, for experimental reason, we can interpret as a speed.
Hence, we look for a linear stochastic dynamics in two dimension and we would like to distinguish between the following two cases: always in equilibrium (I) and generally not (II).
\begin{widetext}
\bea
(\text{I}) \lx\{
\begin{array}{l}
    m\dot{x}_1 + \eta x_1 + k x_2 = \xi \\
    \dot{x}_2 = x_1 \\
    \ave{\xi(t)\xi(t')}=2T\delta(t-t')
\end{array} \rx.
(\text{II})
\lx\{
\begin{array}{l}
    \dot{x}_1 + \alpha x_1 - \lambda x_2 = \xi_1 \\
    \dot{x}_2 -\mu x_1 + \gamma x_2 = \xi_2 \\
    \ave{\xi_i(t)\xi_j(t')}=\nu_{ij}\delta(t-t')
\end{array} \rx.
\text{  with  }
\lx\{
\begin{array}{lcl}
\eta>0, \;\; k>0 \\
l_+ + l_- = \eta / m = \alpha+\gamma = \cT \\
l_+  l_- = k/m = \alpha\gamma-\lambda\mu = \cD \\
\cLfo = (i\om)^2 + i\om \cT + \cD \\
\nu_{ij}=\nu_{ji}
\end{array}\rx.
\eea
\end{widetext}
where $l_+,l_-,\cT, \cD$ are the eigenvalues, trace and determinant of $A$ respectively.
A simple computation prove that there is a substantial difference in the Fourier transform of correlation function $\cCf_y(\om)$. This difference allow us to distinguish the two cases simply by looking at some conditions on the coefficients $c_+$ and $c_-$.
\begin{widetext}
\bea
\cCf_y(\om) = \frac{1}{\lx|\cLfo\rx|^2}
\lx\{
\begin{array}{lcl}
(2T/m) \om^2 = c_1 \om^2 & (\text{I}) \rightarrow &
\lx\{
\begin{array}{l}
     c_+l_- + c_-l_+ = 0   \\
     c_+l_+ + c_-l_- = c_1 \\
\end{array} \rx.\\
( \nu_{11} \gamma^2+2\gamma \lambda \nu_{12}+\nu_{22} \lambda^2 ) + \nu_{11} \om^2 = c_0+c_1\om^2  
& (\text{II}) \rightarrow &
\lx\{
\begin{array}{l}
     c_+l_- + c_-l_+ = c_0 /\cD \neq 0 \\
     c_+l_+ + c_-l_- = c_1  \\
\end{array} \rx. \\
\end{array} \rx.
\eea
\end{widetext}
Note that, since matrix $\nu$ is positive definite, the two coefficients $c_0$ and $c_1$ can never be negative, then, if matrices $A$ and $B$ are strictly positive we are sure we are not in case I. But, 
what happens when we have excluded case I? Is it possible to discriminate equilibrium from non-equilibrium in case II? Unfortunately, there are an infinite number of processes like (II) which 
have exactly the same correlation function $\cCf_y(\om)$ of an out-of-equilibrium process while satisfing equilibrium condition $(\alpha-\gamma)\nu_{12}=\lambda \nu_{22}-\mu \nu_{11}$.
For example, given $\cD,\cT, c_0 \text{ and }c_1$ which completely characterize the correlation function, we can look at the equilibrium processes just by fixing $\nu_{12}=0$ and by choosing the parameters as
\bea
\label{eq:equivalent_process}
\gamma = \lx(\frac{c_0}{c_1} + \cD\rx)/\cT \qquad \alpha = \cT -\gamma \\
\lambda \mu = \frac{c_0}{c_1}-\gamma^2 \qquad \frac{\lambda}{\mu} = \frac{c_1}{\nu_{22}}
\nu_{11}  = c_1
\eea
and we are still free to choose any value for $\nu_{22}$. In other words, we are not able to detect the temperatures of the thermal baths since they are mixed with the deterministic forces in the relative residues.
Definitely, it seems impossible understand from the simple knowledge of the correlation function $\cC_y(t)$ of the single dynamical variable $y$ if the original multidimensional system was out or in equilibrium.
\section{A way out: “a posteriori” response}
\label{wayout}
Let us consider an experiment from which we are able to get the time series of a single scalar observable for which the single-time fluctuations $y$ around the average value (that we assume we are able to subtract step-by-step) are, at least a first approximation, normally distributed. 
We estimate the time correlation function $\cC_y(t)$ from the time series of such fluctuations $y$ and then, with in mind the idea of a underlying multidimensional Ornstein–Uhlenbeck process, we fit $\cC_y(t)$ with a linear combination of exponentials. Let's imagine now that we have a knob that, even if in an uncontrolled way, slightly modifies the parameters with which the system is evolving. How will the relative correlation function be made?
If the knob has changed the drift parameters then we will find something almost completely different as the poles will certainly have moved. But if the poles have not moved, we can think that we have changed only the "temperatures" $T_i\to T_i'$ of the effective thermal baths with which we can characterize the contribution of the noise to the process. If so, only the coefficients have changed $c_\alpha \to c_\alpha'$ accordingly with the formula $c_\alpha = \sum_i T_i c_{i\alpha}$ where the $c_{i\alpha}$ depend on the drift only (see Appendix B). Still in the appendix B, we show that, when the system is at equilibrium, all the temperatures $T_i$ are proportional $T_i = T g_i$ to the a  single temperature $T$ with constants of proportionality $g_i$ which depend on the drift only:  in other word, at equilibrium we have a single effective thermal bath.
In light of this observation, after the knob has changed only the noise properties, two things can happen: either the $c_\alpha$ all scale by the same factor or not.
If they scale by the same factor there are two possible explanation:
\begin{itemize}
    \item there is a single thermal bath and the system, both before and after turning the knob, was in equilibrium but at two different temperatures, the ratio between these two temperatures coincides with the ratio between the coefficients $c_\alpha$ of the correlation functions ;
    \item the system is out-of equilibrium but the knob, for some reason, increases or decreases all the effective temperatures of the thermal baths by the same factor $T_i \to r T_i$.
\end{itemize}
The idea of having a knob that varies some parameters of the model has already been used to infer the thermodynamic properties of a system. For example, in~\cite{polettini2017effective} the authors suppose that they can vary the probability of a link in a Markov chain in order to find the stalling condition (no current through the link). This condition is then used to provide a lower bound of the entropy production of the system. Moreover, in the experiments described in~\cite{ciliberto2013heat,ciliberto2013statistical} the experimenters had a knob that allowed to change the temperature of one of the two thermal baths. 
The procedure described above allows us to understand, through two system measurements, whether both measurements were made in equilibrium conditions or not. However, in the second case, it does not allow to make precise statements on the two measures individually, that is, it does not allow to distinguish the three different situations:
\begin{itemize}
    \item the system is in equilibrium and the knob takes it out-of equilibrium,
    \item the system is out-of equilibrium and the knob takes it in equilibrium,
    \item the system is out-of equilibrium both before and after turning the knob.
\end{itemize}
Since the $c_\alpha=\sum_i T_i c_{i\alpha}$ are written as a linear combination of the temperatures $T_i$ and the relative coefficients $c_{i\alpha}$ are known functions of the parameters of the model, it could be possible to distinguish these three cases by taking different measurements by varying several times the effective temperatures of some thermal baths.
Let $D$ be the number of poles of the $\cCfo$ and $n$ the number of the effective thermal baths whose temperatures change by moving our knob.
In this way the relationship between the $c_\alpha$ and the temperatures of the thermal baths is $c_\alpha = q_\alpha + \sum_{i=1}^n T_i c_{i \alpha}$
where the $q_\alpha$ is a $D-$dimensional constant vector which does not depend on the temperatures of the thermal bath we change with the knob and the $c_{i\alpha}$ are $n D$ coefficients with depend on the drift only.
At each measurement we have $D$ conditions but also $n$ additional unknowns temperatures so, with $m$ measurements, the number of the unknown parameters are $(n + 1) D + m n$ and the number of conditions that must be satisfied at the same time is $D m$.
This means that we need at least $m\geq (n+1)D/(D-n)$ measurements in order to be able to fit the $q_\alpha$ and the $c_{i\alpha}$.
Note that the knowledge of the poles entails other $D$ additional conditions that the parameters of the model must satisfy. Hence,
in some particular experimental setups, by putting all these information together we could be able to infer the nature in or out equilibrium from few measurements, as we will show in the next section.
\subsection{Our Protocol at Work for the Brownian Gyrator}
\begin{figure*}[htp]
\centerline{\includegraphics[scale=0.5]{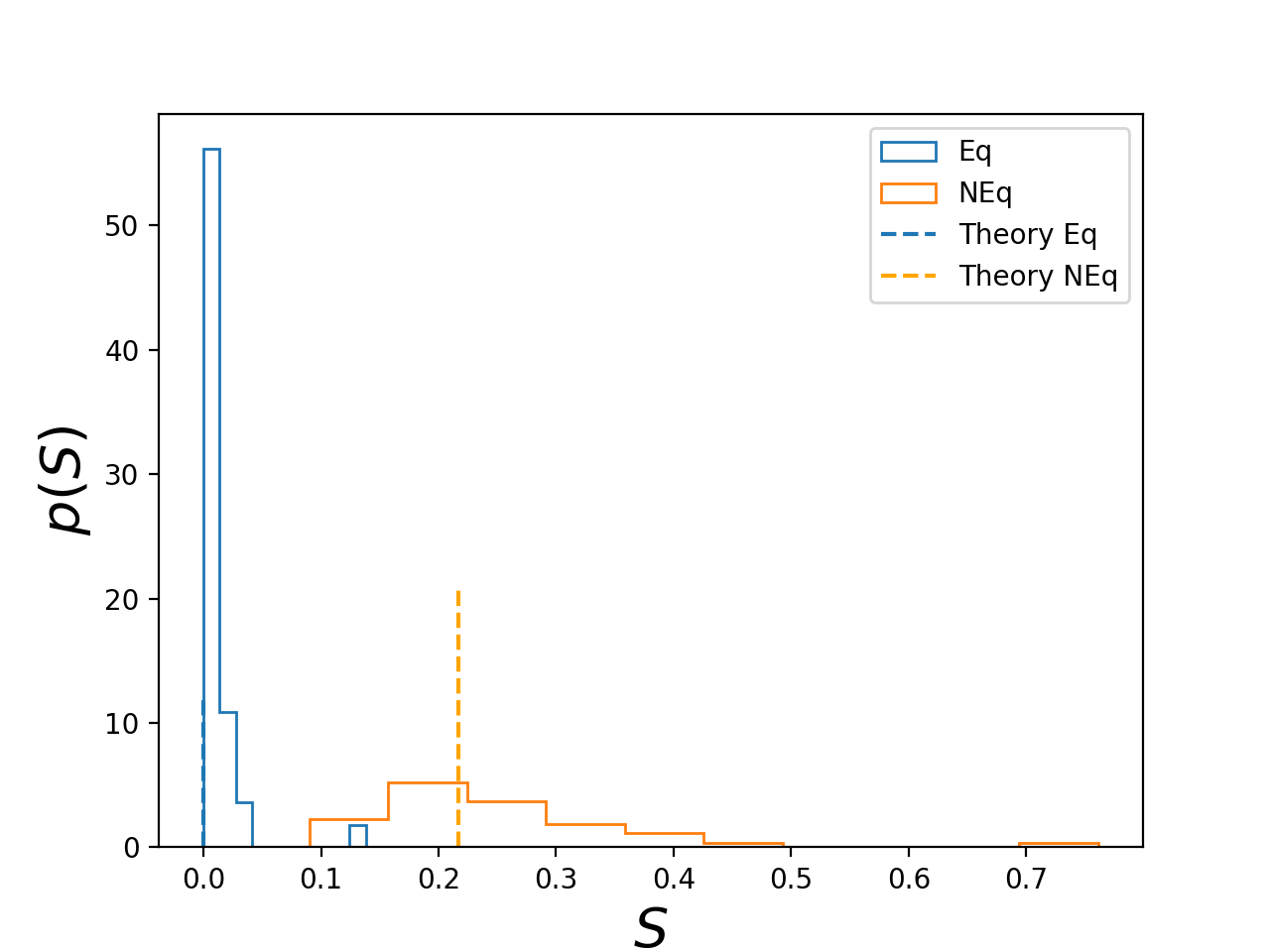}\includegraphics[scale=0.5]{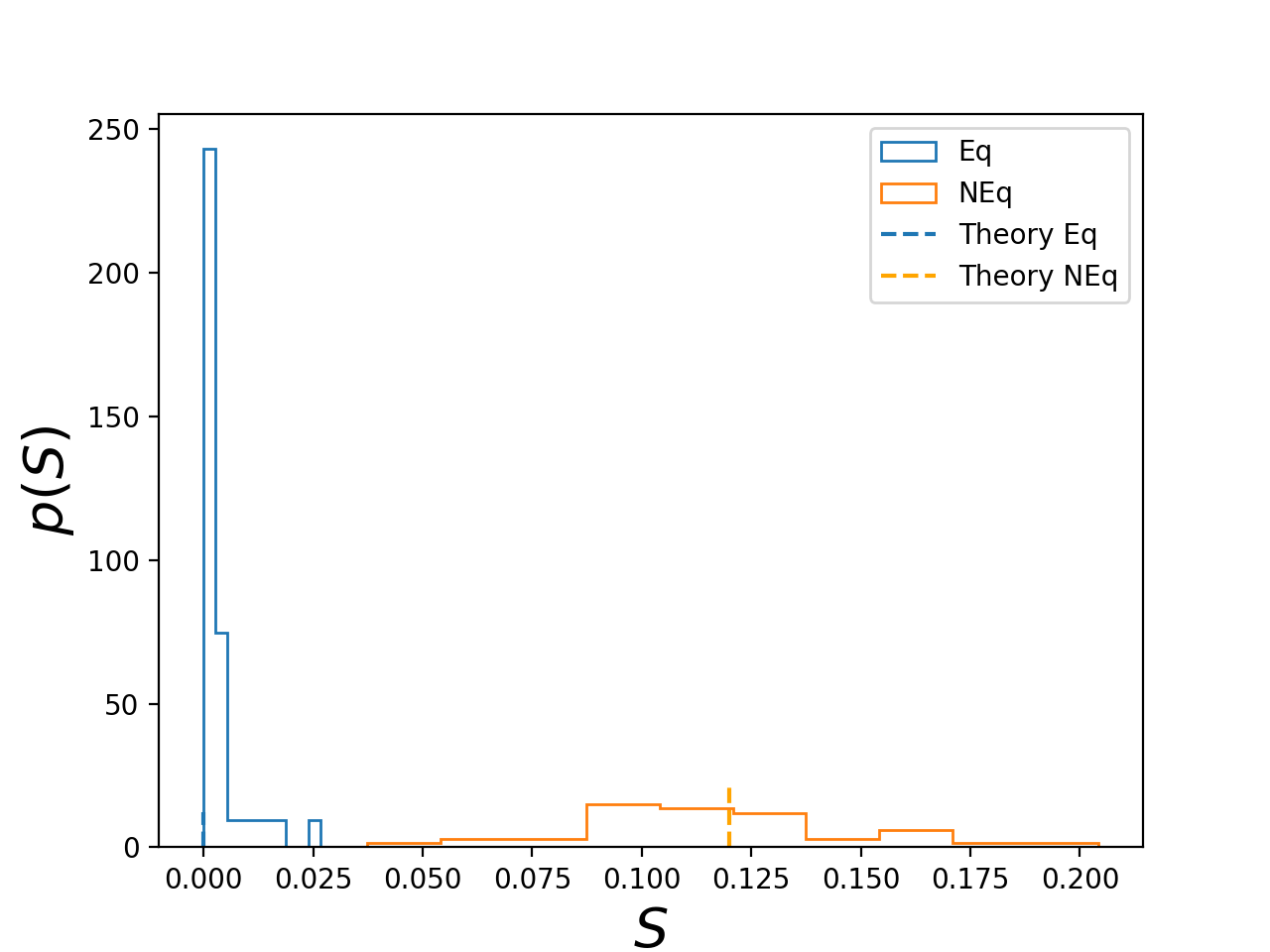}}
\caption{Distributions of the average entropy production rate $S$ for the two systems $S_1$ (left) and  $S_2$ (right). The dotted lines correspond to the theoretical value of the average entropy production rate $S$.}
\label{fig:Entropy_production_histograms}
\end{figure*}
Consider again the example of Sec.\ref{examples}. In this case, the correlation function takes the form $$\cC_y(t)=c_+e^{-l_+|t|} + c_- e^{-l_- |t|}$$ with
\begin{align*}
  c_+ &= \frac{T_1 l_+ \mathcal{D}-l_- \mathcal{Q}}{\mathcal{T}\mathcal{D}(l_+-l_-)}\\
  c_- &= \frac{l_+\mathcal{Q}-T_1 l_- \mathcal{D}}{\mathcal{T}\mathcal{D}(l_+-l_-)}\\
  \mathcal{Q}&=T_1 \gamma^2+T_2 \lambda^2,
\end{align*}
where $\mathcal{T}=\alpha+\gamma$ and $\mathcal{D}=\alpha\gamma-\lambda\mu$. From the definition of $c_+$, $c_-$, $l_+$ and $l_-$ we get
\begin{align}
&l_+c_++l_-c_-=T_1, \nonumber\\
&l_+l_-(l_-c_++l_+c_-)=T_1\gamma^2+T_2\lambda^2.
\label{eq:parameters_estimate_special}
\end{align}
Let $r_1^{(j)}=(l_+c^{(j)}_++l_-c^{(j)}_-)$ and $r_2^{(j)} =l_+l_-(l_-c^{(j)}_++l_+c^{(j)}_-)$ where the index $j$ refers to the $j$-th experiment and imagine that the knob changes only the temperature $T_1$. 
Then, we have
\begin{align}
&\gamma^2=\frac{r^{(1)}_2-r^{(2)}_2}{r^{(1)}_1-r^{(2)}_1},\nonumber\\
&T_2\lambda^2=\frac{r^{(1)}_2\left(r^{(1)}_1-r^{(2)}_1\right)-r^{(1)}_1\left(r^{(1)}_2-r^{(2)}_2\right)}{r^{(1)}_1-r^{(2)}_1}. 
\end{align}
From the knowledge of $\gamma$, we can estimate $\alpha=l_++l_--\gamma$ and $\lambda\mu=\alpha\gamma-l_+l_-$. The entropy production $S$ is proportional to $T_1\mu-T_2\lambda$, i.e.
 \begin{equation}
S = \frac{\left(T_1\mu-T_2\lambda\right)^2}{(l_++l_-)T_1T_2}=\frac{\left(T_1\mu\lambda-T_2\lambda^2\right)^2}{(l_++l_-)T_1T_2\lambda^2}.  
\label{eq:PE}
\end{equation}
\begin{table}[h!]
\centering
\begin{tabular}{|c|c|c|c|c|}
%\hline
%\multicolumn{4}{|c|}{}\\
\cline{2-5}
\multicolumn{1}{c|}{} & \multicolumn{4}{|c|}{\textbf{\Large Entropy production rate}}\\
%\multicolumn{4}{|c|}{}\\
%\hline
\cline{2-5}
\multicolumn{1}{c|}{} & \multicolumn{2}{|c|}{\textbf{System $1$}} & \multicolumn{2}{|c|}{\textbf{System $2$}}\\
%\hline
\cline{2-5}
\multicolumn{1}{c|}{} & \multicolumn{1}{|c|}{\textbf{Theory}} & \textbf{Experiment} & \textbf{Theory} &
 \multicolumn{1}{c|}{\textbf{Experiment}}\\
\hline
\multicolumn{1}{|c|}{Eq} & $0.$ & $0.01\pm0.01$ & $0.$ & $0.003\pm0.003$ \\
\hline
\multicolumn{1}{|c|}{NEq} & $0.217$ & $0.25\pm0.05$ & $0.12$ & $0.115\pm 0.015$\\
\hline
\end{tabular}
\caption{Comparison between theoretical and experimental entropy production rate for the two systems.}
\label{table:1}
\end{table}
Since two measures are sufficient to compute the right hand side of Eq.\ref{eq:PE}, we are able to infer whether the system is at equilibrium or not and furthermore we estimate the average Lebowitz-Spohn entropy production rate $S$.
While the above procedure is theoretically correct, to be useful it must also work in practical cases. We therefore decided to apply this procedure to the two systems introduced in Sec.\ref{examples}. For both systems, different trajectories were simulated by employing the algorithm described in Appendix C. These data were used to estimate the correlation functions and the 4 parameters $l_+$, $l_-$, $c_+$ and $c_-$. Then, the temperature $T_1$ of the thermal bath coupled to $y=x_1$ was changed (system $1$ was put out of equilibrium while system $2$ at equilibrium) and new data were generated. By re-estimating the correlation functions and combining the new measurements with the previous ones, we were therefore able to compute the entropy production $S$ of each of the two systems before and after the manipulation. To check the robustness of the procedure and also to get an idea of the error associated with the estimate of $S$, $40$ experiments were repeated on each system. The results of these experiments are summarized in Fig.\ref{fig:Entropy_production_histograms} which shows the histograms of the entropy production for the two systems in the two cases. 
As can be seen from the histograms, there is a clear difference between equilibrium and non-equilibrium cases. In fact, in the first case the histograms have a peak around zero while in the second case the distributions are wider and have a maximum for non-zero values. Furthermore, it should be noted that in non equilibrium cases the maximum of the distributions is close to the theoretical entropy production. 
Tab.\ref{table:1} shows the theoretical values of the entropy production rate as well as the results obtained experimentally which are compatible considering the error bars\footnote{Note that the errors in Tab.\ref{table:1} are taken to be $3$ standard deviations of the means}.
%%%%%%%%%%%%%%%%%%%%%%%%%%%%%%%%
\section{Coarse Graining of Markov chains}
\label{markov}
\begin{figure*}[htp]
\subfloat[][\emph{} ]
	{\includegraphics[height=.4\textwidth,width=.45\textwidth]{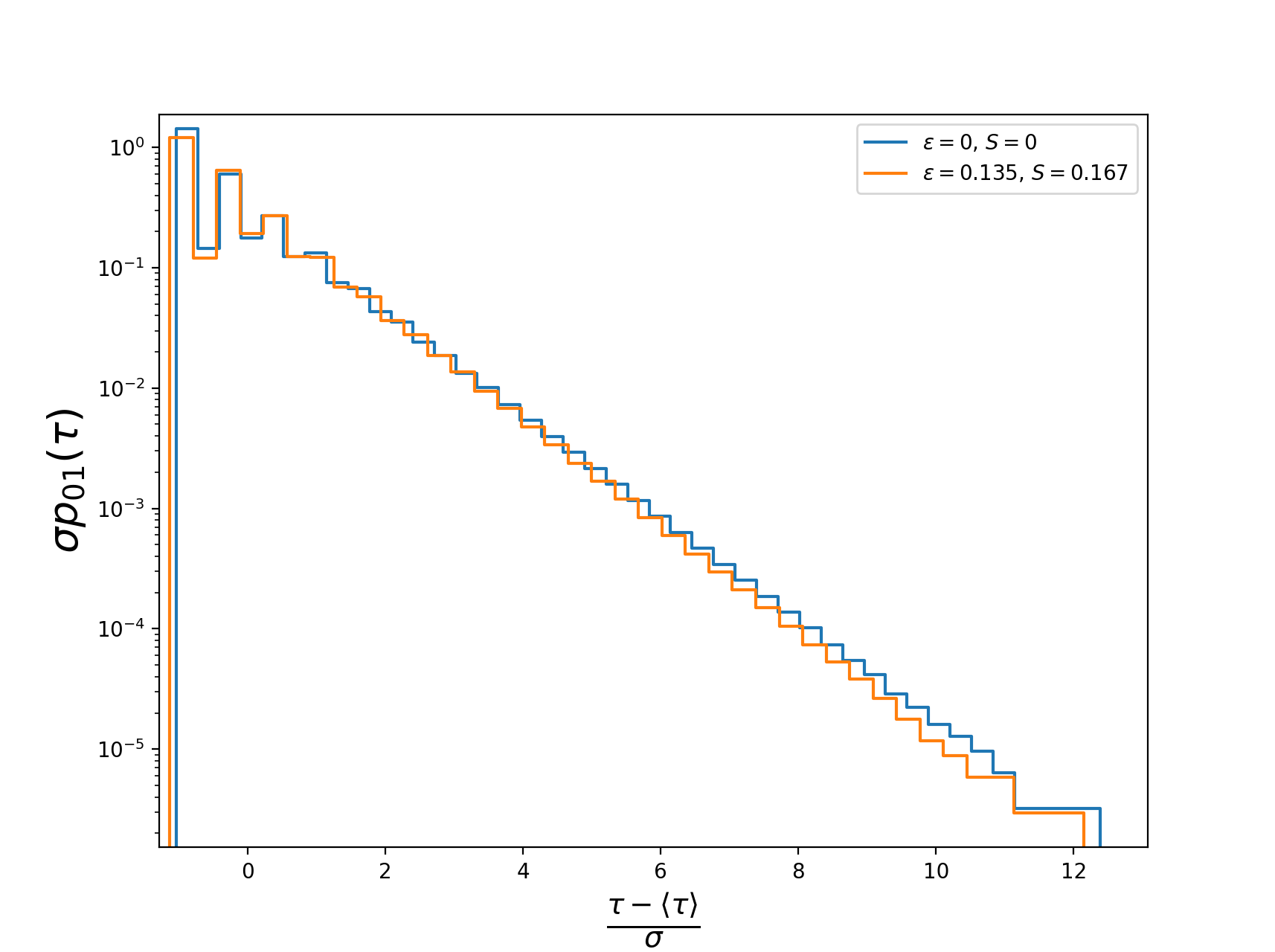} }\quad
\subfloat[][\emph{}]
	{\includegraphics[height=.4\textwidth,width=.45\textwidth]{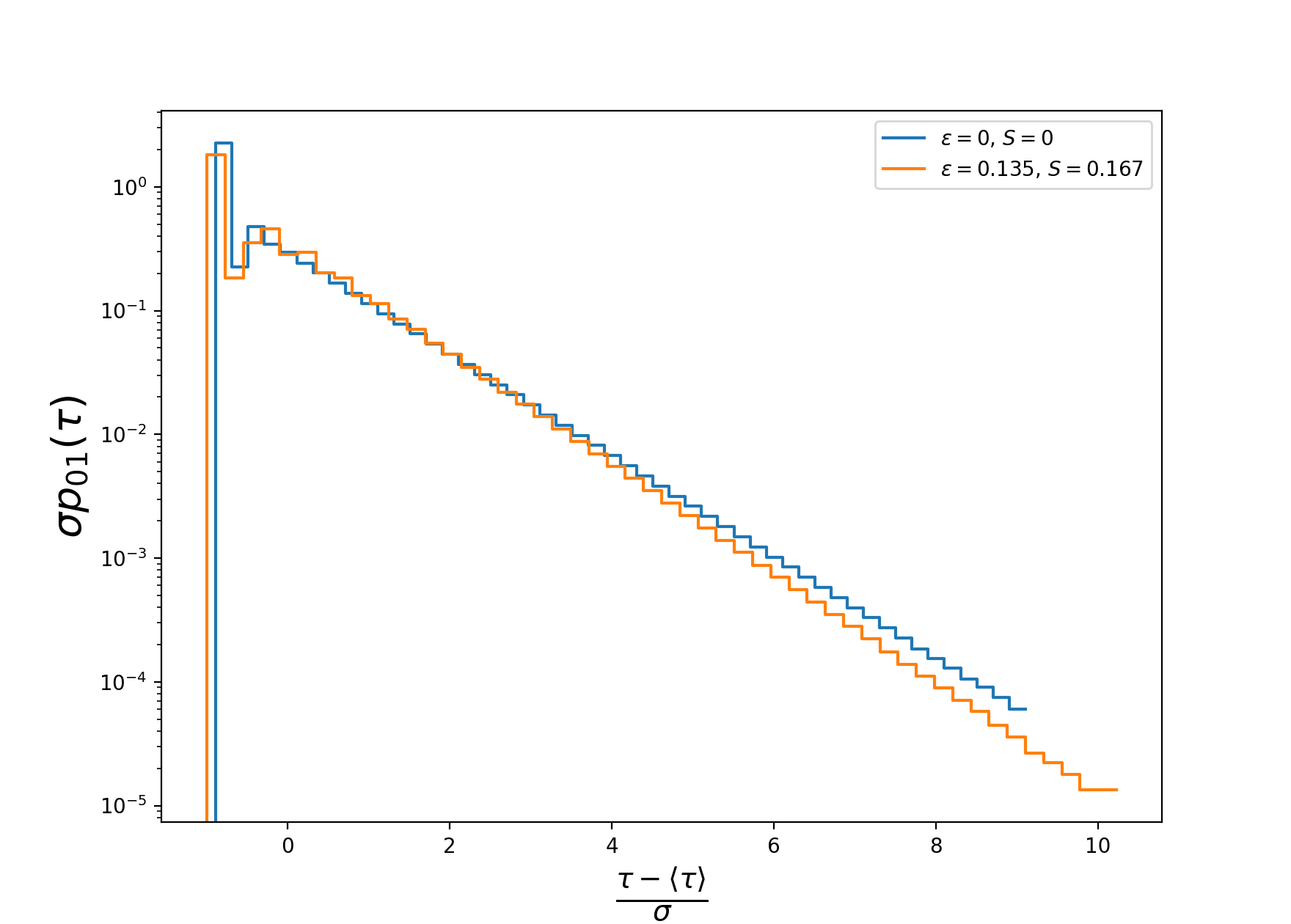}}\\
\subfloat[][\emph{} ]
	{\includegraphics[height=.4\textwidth,width=.45\textwidth]{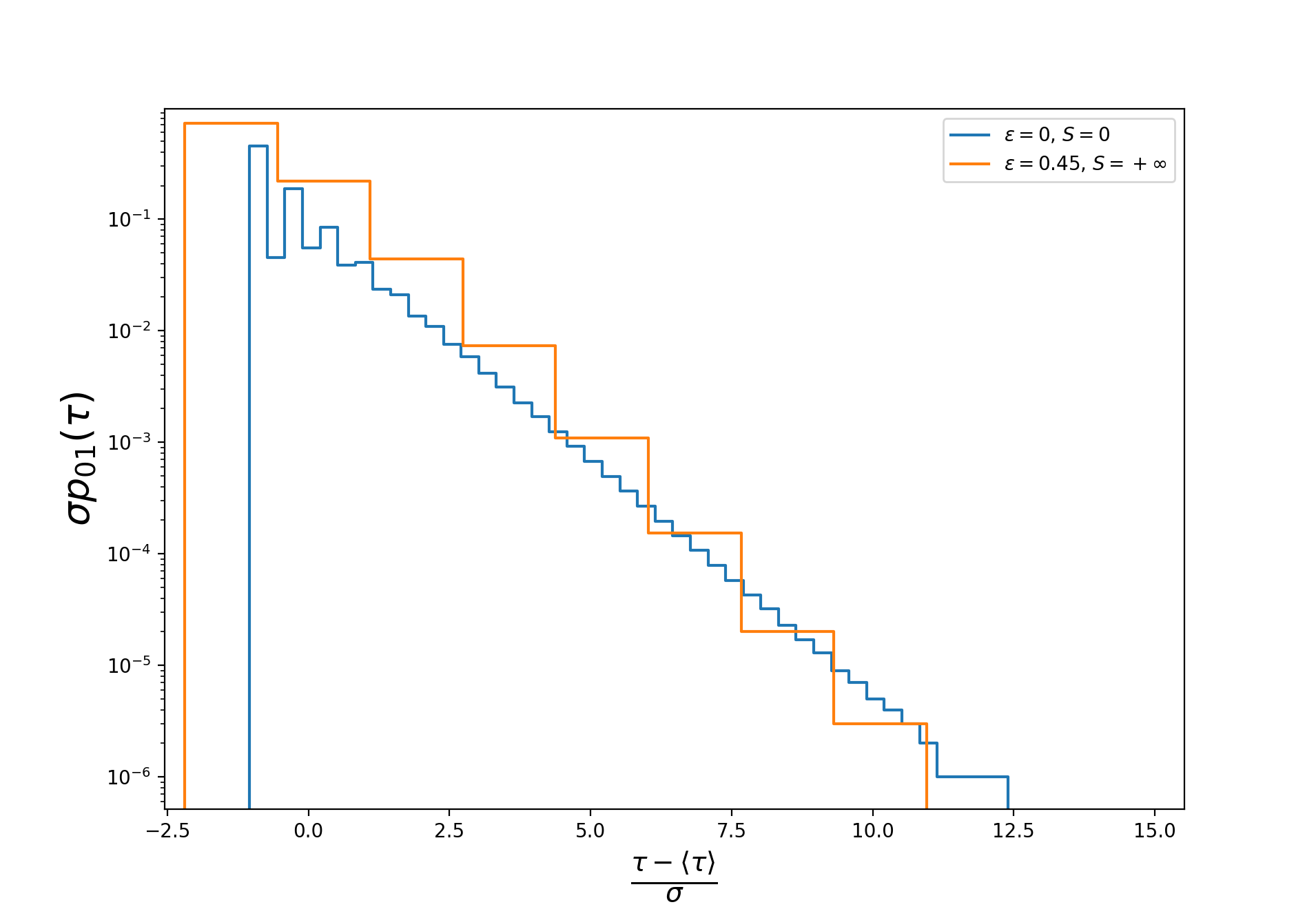} }\quad
\subfloat[][\emph{}]
	{\includegraphics[height=.4\textwidth,width=.45\textwidth]{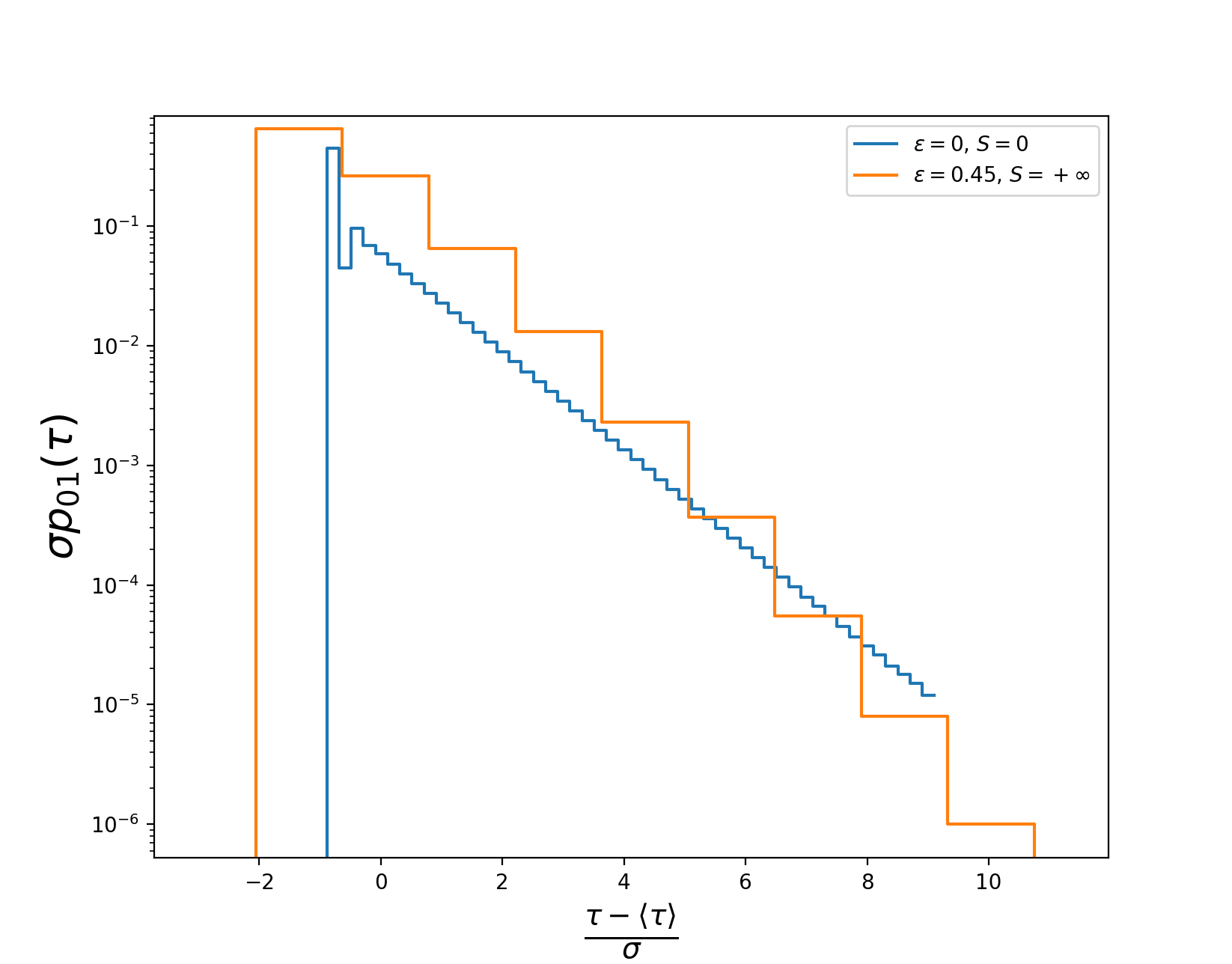}}\\
\caption{Distributions of rescaled exit times $\tau_r$ from state $0$ to state $1$ for Markov chains on a ring. Left panels (a-c): Markov chain with $N=4$ states. Right panels (b-d): Markov chain with $N=5$ states. Top panels (a-b): $\epsilon=0.135$ and $S=0.167$. Bottom panels (c-d): $\epsilon=\frac{1-\alpha}{2}=0.45$ and $S=+\infty$.}
\label{fig:ExitTimes}
\end{figure*}
So far we have only dealt with linear stochastic processes whose states are represented by vectors in $\mathbb{R}^D$. However, it may happen that an appropriate description of the problem requires the use of a discrete phase space. In these cases, the state of the system is represented by an integer index $i=1, 2 \cdots,N$. In analogy with continuous processes, the experimenter usually does not have access to all the phase space and is therefore limited to observe a coarse-grained process that lives in a reduced phase space. In this Section we show that, in the context of Markov chains or Markov jump processes, some coarse-graining procedures lead to results similar to those of Gaussian continuous systems, i.e. that the statistical features of the coarse-grained process and of its time-reversal are indistinguishable despite the underlying process being an out of equilibrium Markov process. In the following, we discuss the case of Markov chains but the results are correct also for processes with continuous time.

Let $\Omega=\{1, 2 \cdots,N\}$ be the phase space of a system described by a Markov chain whose transition matrix is denoted by $\Pab$. Now imagine that an experimenter is not able to observe all the states of the system but rather he observes a coarse-grained process where the states have been grouped into two disjoint groups. Let $a=0,1$ represents the state of the coarse-grained process. Given the nature of the problem, it is natural to introduce a block representation of both $\Pab$ and the invariant distribution $\pa$, i.e.
\begin{widetext}
\begin{eqnarray*}
\pa &=& \lx( 
\begin{array}{c|c}
    \begin{array}{ccc}
    \pi_1 & \cdots & \pi_m \\
    \end{array} & 
    \begin{array}{ccc}
    \pi_{m+1} & \cdots & \pi_N \\
    \end{array}
\end{array} \rx) = 
\lx( 
\begin{array}{c|c}
\pa_0 & \pa_1 \\
\end{array} \rx) \, ,\\
\Pab &=& 
\lx( 
\begin{array}{c|c}
    \begin{array}{ccc}
    G_{1,1} & \cdots & G_{1,m} \\
    \vdots & \ddots & \vdots \\
    G_{m,1} & \cdots & G_{m,m} \\
    \end{array} & 
    \begin{array}{ccc}
    G_{1,m+1} & \cdots & G_{1,N} \\
    \vdots & \ddots & \vdots \\
    G_{m,m+1} & \cdots & G_{m,N} \\
    \end{array} \\
    & \\
    \hline \\
    \begin{array}{ccc}
    G_{m+1,1} & \cdots & G_{m+1,m} \\
    \vdots & \ddots & \vdots \\
    G_{N,1} & \cdots & G_{N,m} \\
    \end{array} &
    \begin{array}{ccc}
    G_{m+1,m+1} & \cdots & G_{m+1,N} \\
    \vdots & \ddots & \vdots \\
    G_{K,m+1} & \cdots & G_{N,N} \\
    \end{array} \\
\end{array} \rx) = 
\lx(
\begin{array}{c|c}
\Pab_{00} & \Pab_{01} \\
& \\
\hline \\
\Pab_{10} & \Pab_{11} \\
\end{array} \rx) \, ,\\
&&
\one_0 =
\lx.
\begin{pmatrix}
1 \\ \vdots\\ 1 \end{pmatrix} \rx\} m
\qquad
\one_1 = 
\lx.
\begin{pmatrix}
1 \\ \vdots\\ 1 \end{pmatrix} \rx\} N-m \,.\\
\end{eqnarray*}
\end{widetext}
The probability of a sequence $\bsa=\{a_t\}_{1\le t \le T}$ of length $T$ (as well as the other statistical quantities that characterize the process) can be computed from the knowledge of $\Pab$ and the invariant distribution $\pa$, i.e.
\begin{equation}
 P(\bsa)=\pa_{\sa_1} \prod_{t=1}^{T-1} \Pab_{\sa_{t}\sa_{t+1}} \one_{\sa_T}. 
\end{equation}
Also note that the sequence $\bsa$ can be encoded with a sequence of $K$ pairs $(a_k, n_k)$ where $n_k$ represents the time spent in the macrostate $a_k$ and therefore
\begin{equation}
P(\bsa) = P(a_1,n_1;a_2,n_2;\cdots;a_K,n_K),
\end{equation}
with $\sum_kn_k=T$ and $a_{k+1}=\bar{a}_k\equiv 1- a_k$.
In general, the process describing the evolution of $a$ is non-Markovian and the computation of the KL divergence between the probability of time-forward and time-backward sequences provides a lower bound on the production rate of entropy of the whole system. 

Now consider the special case in which one of the two macrostates, say the macrostate $1$, contains only one microstate. Since the state $1$ is pure and the whole process is Markovian, the dynamics of the coarse-grained process is correlated only in the time-interval between two successive visits of this state. To put it another way, the coarse-grained process is a semi-Markov process. Let $p_{a\bar{a}}(\tau)$ the probability distribution of the exit times from state $a$. Since the process is semi-Markov, we have that the probability of the sequence $\bsa$ is
\begin{equation}
P(\bsa)=p^{in}_{a_1a_2}(n_1)\left(\prod_{k=2}^{K-1}p_{a_k a_{k+1}}(n_k)\right)p^f_{a_K}(n_K), 
\end{equation}
where $p^{in}$ ($p^f$) is the initial (final) probability of observing a sequence starting (ending) with $n_1$ ($n_K$) characters $a_1$ ($a_K$).
The reverse sequence will instead be $\obsa=(a_K,n_K;a_{K-1},n_{K-1};\cdots;a_1,n_1)$ and its probability is
\begin{equation}
P(\obsa)=p^{in}_{a_K,a_{K-1}}(n_K)\left(\prod_{k=2}^{K-1}p_{a_{k}a_{k-1}}(n_k)\right)p^f_{a_1}(n_1). 
\end{equation}
Since $a_{k-1}=a_{k+1}$, $P(\bsa)$ and $P(\obsa)$ differ only for the boundary terms. Therefore, when the length $T$ of the two sequences $\bsa$ and $\obsa$ goes to infinity the two probabilities are equal and the entropy production rate vanishes, i.e.
\begin{equation}
S=\lim_{T+\infty}\frac{1}{T}\sum_{\bsa} P(\bsa)\log\left(\frac{P(\bsa)}{P(\obsa)}\right)=0. 
\end{equation}

This result seems to leave no hope of understanding the thermodynamic state of the system through the observation of the coarse-grained process. However, in~\cite{skinner21b} the authors show that even in these cases there is a more powerful analysis based upon residence time statistics, showing - in some cases - that the only Markovian processes compatible with the observations are not at equilibrium. Note that the same is true for linear processes whose correlations have sinusoidal modulations (see the discussion in appendix B.3). 

Nevertheless, we expect that, in general, for such a coarse-grained procedure it will be possible to find two Markov chains, one at equilibrium and one out of equilibrium, that produce the same statistics for the exit times. 

To support this conjecture we have considered Markov chains with a simple topology, that is, translation invariant Markov chains on a ring with periodic boundary condition of size $N$, for which the invariant distribution is uniform and the transition matrix is such that
\bea
G_{ii}=\alpha \quad G_{ii+1}=\frac{1-\alpha}{2}+\epsilon  \quad G_{ii-1}=\frac{1-\alpha}{2}-\epsilon
\eea
and $G_{ij}=0$ otherwise.
In Fig.\ref{fig:ExitTimes}, we show the comparison between the distributions of the rescaled exit times, i.e. $\tau_r=\frac{\tau-\langle\tau\rangle} {\sigma}$, in the cases at equilibrium ($\epsilon = 0 $) and out of equilibrium ($ \epsilon \neq 0 $) for two different chains with $ N = 4 $ (left) and $ N = 5 $ (right) states.
As might be expected, for small values of $ \epsilon $ (top panels) it is very difficult to distinguish a distribution that originates from an equilibrium process from one determined from an out of equilibrium process. More surprisingly, also in the case with $G_{ii-1}=0$ (completely irreversible process) the distribution of the rescaled exit times is not too dissimilar from its equilibrium counterpart (see bottom panels).
%%%%%%%%%%%%%%%%%%%%%%%%%%%%%%%%
\section{Conclusions}
\label{conclusions}

In this paper we have shown results whose mathematical aspects  are in part already present in the (physical or mathematical) literature~\cite{weiss1975time,diks1995reversibility,zamponi2005fluctuation,crisanti2012nonequilibrium}, quite scattered in time and  not widely known, i.e. that one-dimensional Gaussian data - even when non-Markovian and coming from a system which is out-of-thermodynamic equilibrium - are always indistinguishable from equilibrium. In all the papers where we have found something of it, the Authors do not draw conclusions about the inference problem neither they propose strategies to circumvent the observed obstacles: in the present paper we do both. 

After discussing the problem in its full generality, we have given concrete examples where $1d$ data coming from equilibrium and non-equilibrium systems are indistinguishable, even if an observation in full phase space shows very strong differences. This result appears more surprising when looking to quantities such as bridges which are strongly asymmetric in the full phase-space  of an out-of-equilibrium system, they are still asymmetric if the bridge observed in full phase-space is projected in $1d$, but lose completely their asymmetry when they are constructed directly in the reduced ($1d$) space. We have discussed a general demonstration of the problem, which amounts to the fact that correlations, which are the only information contained in Gaussian-distributed data, contains an entangled product of quantities related to both memory and noise. Disentangling these two ingredients would allow us to check the 2nd-kind fluctuation-dissipation relation~\cite{kubo2012statistical}, but in $1d$ this is indeed impossible. An additional conclusion that can be drawn from this observation is that linear response cannot be deduced, in general, from correlations, in a $1d$ non-Markovian systems. We underline that these negative results bear a certain degree of surprise if one expects  analogy with deterministic systems to hold. In deterministic systems, in fact, the reconstruction of all the properties of a system in dimensions larger than $1$ can be done starting from a (long enough) time-series of a $1d$ observable (embedding thecnique~\cite{takens1981detecting,vulpiani2009chaos}). Our discussion, therefore, is a more convincing proof that for stochastic systems the embedding idea is not going to work, in general.

\appendix
%%%%%%%%%%%%%%%%%%%%%%%%%%%%%%%%
\section{Entropy production}

Here we briefly recall the fact that a partial observation cannot overestimate the EP of a system, being it linear, non-linear, Markovian or not Markovian.
We first wish to show an example of how taking longer - but incomplete - observations can improve our knowledge of a deterministic system (approaching, for large times, the knowledge of the full phase-space, as in the embedding Takens theorem) while this is not true for stochastic systems. Consider a  deterministic process in discrete time, so that at time $i$ it stays in state $\gamma_i$. When the time goes from $1$ to $t$ the process generates the path $\Gamma_t=\gamma_1...\gamma_t$ which is fully determined by initial state $\gamma_1$.  One observes the system without maximum precision, that means that instead of the path $\Gamma_t$ the observer sees a path $\Omega_t=\omega_1...\omega_t$: the lack of precision is in the fact that many different $\Gamma_t$ correspond to the same observed $\Omega_t$, we can define the set $C(\Omega_t)$ which contains all the paths $\Gamma_t$ compatible with the observation $\Omega_t$. When a new $\omega_{t+1}$ state is observed the number of compatible $\Gamma_{t+1}$ can remain the same or reduce, it cannot grow because the new observation is an additional constraint on a fixed information (the initial state $\gamma_1$). So, observing longer and longer paths $\Omega_t \to \Omega_{t+1} \to \Omega_{t+2}$ implies smaller and smaller sets of compatible states $C(\Omega_t) \supseteq	C(\Omega_{t+1})  \supseteq	C(\Omega_{t+2})  \supseteq...$. This suggests that a long enough time-series of a partial observation should be equivalent to the observation of the system in full phase space increasing the length $t$ of the observed path $\Omega_t$. 
Now imagine to repeat the reasoning above for a stochastic process. Each new observation  $\omega_{t+1}$ has not the same power as in the deterministic case: in fact the underlying new state $\gamma_{t+1}$ is not exactly determined by the previous story, therefore $\omega_{t+1}$ is an additional constraint on a string $\Gamma_{t+1}$ which also contains (in general) more information than $\Gamma_t$: there is no reason to expect  $C(\Omega_t) \supseteq	C(\Omega_{t+1}) $ and therefore it is not true, in general, that a longer partial observation can help in inferring features of the full phase space. This fact has been observed for continuous systems in continuous time in~\cite{baldovin2020understanding}.

The lack of information about the full phase space affects entropy production in the following way. By defining $\Gamma_t^*$ the time-reversed path, the average entropy production measured in a time-length $t$ is
\begin{equation}
S^\Gamma_t=\sum_{\Gamma_t} P(\Gamma_t) \ln \frac{P(\Gamma_t)}{P(\Gamma_t^*)},
\end{equation}
when we observe the system with lower precision we can only measure
\begin{widetext}
\begin{equation}
S^\Omega_t=\sum_{\Gamma_t} P(\Gamma_t) \ln \frac{P(\Omega_t[\Gamma_t])}{P(\Omega_t[\Gamma_t]^*)}=\sum_{\Gamma_t} P(\Gamma_t) \ln \frac{\sum_{\Gamma_t \in C(\Omega_t[\Gamma_t])}P(\Gamma_t)}{\sum_{\Gamma_t \in C(\Omega_t[\Gamma_t]^*)} P(\Gamma_t)}.
\end{equation}
\end{widetext}
where the notation $\Omega_t[\Gamma_t]$ the coarse-grained path $\Omega_t$ corresponding to the real path $\Gamma_t$.
So in general $S^\Omega_t \neq S^\Gamma_t$. Most importantly (and perhaps not noticed before), in view of the previous considerations, there is apparently no reason to expect - for stochastic systems - that increasing $t$ may let $S^\Omega_t \to S^\Gamma_t$.

In~\cite{crisanti2012nonequilibrium} (Sec. IIIC) - a simple demonstration is given, for a particular kind of coarse-graining (from 2d to 1d), for continuous stochastic processes - that $S^\Omega_t \le S^\Gamma_t$. The passages can be generalised:
\begin{equation}
S^\Gamma_t-S^\Omega_t=\sum_{\Gamma_t} P(\Gamma_t) \ln \frac{P(\Gamma_t)}{Q(\Gamma_t)} =D_{KL}(P || Q) \ge 0.
\end{equation}
The validity of  the interpretation as a Kullback-Leibler divergence (which is non-negative) is guaranteed by the fact that $Q(\Gamma_t)=P(\Gamma_t^*)P(\Omega_t[\Gamma_t])/P(\Omega_t(\Gamma_t)^*)$ is positive and normalised:
\begin{widetext}
\begin{equation}
\sum_{\Gamma_t} Q(\Gamma_t)=\sum_{\Omega_t} \sum_{\Gamma_t \in C(\Omega_t)} Q(\Gamma_t)=\sum_{\Omega_t} P(\Omega_t) \sum_{\Gamma_t \in C(\Omega_t)}\frac{P(\Gamma_t^*)}{P(\Omega_t(\Gamma_t)^*)}=1.
\end{equation}
\end{widetext}
The last passage requires that $C(\Omega_t[\Gamma_t]^*) \equiv C(\Omega_t[\Gamma_t^*])$~\footnote{This seems reasonable but perhaps can be violated by those coarse-graining procedures that depend on the history.}. Note that $S^\Gamma_t=S^\Omega_t$ if $P(\Gamma_t|\Omega_t)=P(\Gamma_t^*|\Omega_t^*)$ where we have defined $P(\Gamma_t|\Omega_t)=P(\Gamma_t)/P(C(\Omega_t[\Gamma_t])$.

%%%%%%%%%%%%%%%%%%%%%%%%%%%%%%%%
\section{Linear systems}
In this Appendix we give a complete treatment of linear systems of integro-differential stochastic equations with correlated-in-time noise, that should cover the largest possible set of stochastic (Markovian and non-Markovian) processes with Gaussian statistics. Of course the topics is largely treated in the literature, in probability and in physics, starting from the seminal paper of Uhlenbeck and Ornstein in 1930~\cite{uhlenbeck1930theory}, to modern books on stochastic proccesses such as~\cite{gardiner2009stochastic} and~\cite{risken1996fokker} which treat in details the consequences of response theory, time-dependent problem and of the consequences of detailed balance, focusing on the Markovian case. The non-Markovian case is much less a textbook case and for this reason we decided to review the topics in a compact and general way.

\subsection{Correlation and response}
We consider the vector process $x(t)$ that obeys the following equation
\begin{equation}
(\cL x)(t)=(\cB\xi)(t)+h(t)
\end{equation}
where $x(t)=\lx\{ x_i(t) \rx\}_{i=1,D} \in \mathbb{R}^D$ is a set of dynamical variables, $h(t)=\lx\{h_i(t)\rx\}_{i=1,D} \in \mathbb{R}^D$ is a set of  external fields and $\xi(t)=\lx\{\xi_\alpha(t)\rx\}_{\alpha=1,d} \in \mathbb{R}^d$ a vector of colored normally distributed random noise with zero mean $\ave{\xi_\alpha(t)}=0$ and covariance matrix $\ave{\xi_\alpha(t)\xi_\beta(t')}=\nu_{\alpha\beta}(t-t')$
depending on times $t$ and $t'$ just by the difference $(t-t')$. We have also introduced $\cL = \lx\{\cL_{ij}\rx\}_{i,j=1,D}$ and $\cB = \lx\{\cB_{i\alpha}\rx\}_{i=1,D}^{\alpha=1,d}$ which are two matrices whose elements are linear combinations of operators that act on single variables by multiplying, differentiating or integrating them in a convolution, i.e.
$\lx(\cA f\rx)(t) = \int dt \cA(t-t') f(t')$ where the kernels $\cA(t)$ must decay fast enough to make finite the integral on $t\in(-\infty,+\infty)$~\footnote{We do not pretend mathematical rigour, here. Physically the fast decay of  memory kernels are required to get meaningful stationary states and vanishing of memory of the initial conditions in a finite time.}.

The above equation can also written, component by component,  as
\begin{equation}\label{genlinequ}
\sum_j (\cL_{ij} x_j)(t) = \sum_{\alpha} (\cB_{i\alpha}\xi_\alpha)(t) + h_i(t)
\end{equation}
In this framework we consider only real functions and it is useful to look at Fourier space (the overline is the complex-conjugate)
\bea
f(t) = \int \frac{d\om}{\sqrt{2\pi}} \widetilde{f}(\om)e^{i\om t} \longleftrightarrow \widetilde{f}(\om) = \int \frac{dt}{\sqrt{2\pi}} f(t) e^{-i\om t} \\
\eea
( if $f(t)\in \mathbb{R}$ then $\widetilde{f}(-\omega) =\overline{\widetilde{f}(\omega)}$),
introduce a commutative inner product 
\bea
(f,g) &=& \sum_i \int dt\, f_i(t) g_i(t) =\sum_i \int d\om\, \overline{\widetilde{f}_i(\om)} \widetilde{g}_i(\om) = (g,f) ,
\eea
and define the adjoint $\cA^\dagger$ and the inverse $\cA^{-1}$ (if exist) of an generic operator $\cA$ 
\begin{widetext}
\bea
(f,\cA g) &=& (\cA^\dagger f,g) = \sum_{ij} \int dt \int dt' \, f_i(t) \cA_{ij}(t-t') g_j(t') \rightarrow \cA^\dagger(t) = \cA(-t)^T \quad\lx( \cA^\dagger_{ij}(t) = \cA_{ji}(-t) \rx)\\
&=& \sqrt{2\pi} \sum_{ij} \int d\om \, \overline{\widetilde{f}_i(\om)} \cAf_{ij}(\om) \widetilde{g}_j(\om) \longrightarrow \widetilde{\cA^\dagger}(\om) = \cAfo^\dagger \qquad \lx( \widetilde{\cA^\dagger}_{ij}(\om) =  \cAf_{ji}(-\om) = \overline{\cAf_{ji}(\om)} \rx) \\
(f,\cA \cA^{-1} g) &=& (\cA^\dagger f,\cA^{-1} g) = (f,g) \rightarrow \sum_k \int ds \cA_{ik}(t-s)\cA^{-1}_{kj}(s-t') = \delta_{ij} \delta(t-t') \rightarrow \\
&\rightarrow&
\lx\{
\begin{array}{rcl}
\widetilde{\cA^{-1}}(\om) &=& \lx(2\pi \cAfo\rx)^{-1} \\
\cA^{-1}(t) &=& \int \frac{d\om}{\sqrt{2\pi}}\lx(2\pi \cAf(\om)\rx)^{-1} e^{i\om t} \neq
\lx(\cA(t)\rx)^{-1}  \\
\end{array} \rx. \\
\eea
\end{widetext}
paying attention to the  difference between
$\widetilde{\cA^\dagger}(\om)$ (the Fourier transform of operator $\cA^\dagger(t)$) and $\cAfo^\dagger$ (the transposed-conjugated matrix of $\cAfo$) and between
$\widetilde{\cA^{-1}}(\om)$ (the Fourier transform of $\cA^{-1}(t)$) and $\cAfo^{-1}$ (the inverse matrix of $\cAfo$).\\
Now, we are ready to look at equation (\ref{genlinequ}) in a compact way. If we assume that $x(t)$ is known since $t=-\infty$ and up to $t=+\infty$, in Fourier space it obeys
\bea
\cLfo \xfo = \cBfo \xifo + \widetilde{h}(\om)
\eea
In this way the solution will be
\bea
\xfo &=& \sqrt{2\pi} \lx(\cGfo \xifo + \cRfo \widetilde{h}(\om)\rx) \nonumber\\ 
x_i(t) &=& \sum_\alpha \int  dt' \cG_{i \alpha}(t-t') \xi_{\alpha}(t') +\\ &+&\sum_j \int dt'\cR_{ij}(t-t')h_j(t') \label{genlinsol}
\eea
where
\bea
\cGfo = \cRfo \cBfo \text{  and  } \sqrt{2\pi}\cRfo=\cLfo^{-1}  \\
\eea
so, by computing mean $m_i(t)$ and time-correlations function $\cC_{ij}(t,t')$ directly from the solutions of equation (\ref{genlinsol})
\bea
m_i(t) &=& \ave{x_i(t)} = \sum_j  \int dt' \cR_{ij}(t-t')h_j(t')  
 \\
\cC_{ij}(t,t') &=& \ave{x_i(t) x_j(t')}_c = \ave{ ( x_i(t)-m_i(t))(x_j(t')-m_j(t')} = \\
&=& \sum_{\alpha \beta} \int  ds \int ds' \;
\cG_{i\alpha}(t-s) \nu_{\alpha\beta}(s-s') \cG_{j\beta}(t'-s') = \\
&=& 2\pi \sum_{\alpha \beta} \int \frac{d \om}{\sqrt{2\pi}} \; \cGf_{i\alpha}(\om) \widetilde{\nu}_{\alpha\beta}(\om) \overline{\cGf_{j\beta}(\om)} e^{i\om(t-t')} \\
&=& \cC_{ji}(t'-t) \qquad \lx( \widetilde{\nu}(\om)^\dagger = \widetilde{\nu}(\om)\rx)\\
\eea
we get that the linear response function $\lx. \ave{\partial x_i(t)/\partial h_j(t')}\rx|_{h=0}$ is just $\cR_{ij}(t-t')$ and that the time correlation function $\cCfo$ in Fourier space reads
\bea
\cCfo &=& 2\pi \cRfo \cSgfo \cRfo^\dagger = \cCfo^\dagger
\eea
where
\bea
\cSgfo &=& \cBfo\widetilde{\nu}(\om)\cBfo^\dagger =\cLfo \cCfo \cLfo^\dagger = \cSgfo^\dagger.
\eea
Being involved only linear operators and assuming Gaussian-distributed (or delta-peaked) initial conditions,
every (joint or conditional) probability distribution of a sequence of  $m$ observations $\{x(t_1)=x_1,x(t_2)=x_2,\dots,x(t_m)=x_m\}$ is a multivariate Gaussian. 
It is important to stress that even if $x(t)$ is non-Markovian, when $h=0$, the knowledge of $\cC(t)$ is sufficient to reconstruct all such probabilities and therefore the full path probabilities too.

In particular the joint probability distribution  of $m$ observations, in the case $h=0$, reads
\bea
\cP(x(t_1)&=&x_1,x(t_2)=x_2,\dots,x(t_m)=x_m) = \\ &=&\cN_{\widehat{\cC}}(x_1,x_2,...,x_m)
\eea
where the multivariate Gaussian for a generic vector $z$ in $n$ dimension with covariance matrix $\cA$ is 
\bea
\cN_{\cA}(z_1,\dots,z_n) = \frac{1}{\sqrt{|2 \pi \cA|}}\exp{-\frac{1}{2}\sum_{ij}z_i\cA^{-1}_{ij} z_j}
\eea
and $\widehat{\cC}$ is a matrix $mD \times mD$ composed of blocks of the matrix two-time correlation $\cC$ evaluated at the time-differences between all the observations
\bea
\widehat{\cC} = 
\begin{pmatrix}
\cC(0) & \cC(t_1-t_2) & \dots & \cC(t_1-t_m) \\
\cC(t_2-t_1) & \cC(0) & \dots & \cC(t_2-t_m) \\
\vdots & \vdots & \ddots & \vdots \\
\cC(t_m-t_1) & \cC(t_m-t_2) & \dots & \cC(0) \\
\end{pmatrix} 
\eea
Note that, since $\cC(-t)=\cC(t)^T$ we have $\widehat{\cC}^T= \widehat{\cC}$.
\subsection{Detailed balance condition}

Now we discuss the detailed balance condition (when $h=0$), considering the difference between the  joint probability $\cP(x(t_0)=x_0,x(t_1)=x_1,\dots,x(t_{m-1})=x_{m-1})$ and the joint probability of the "reverse path" $\cP(x(t_0)=Sx_{m-1},x(t_1)=Sx_{m-2},\dots,x(t_{m-1})=Sx_0)$  where $S_{ij}=s_i \delta_{ij}$ takes into account the effect of the time reversal parity of the different components  ($s_i\in\{-1,1\}, S^2=I, S^{-1}=S$, for instance positions have $s_i=1$ and velocities have $s_i=-1$).
Such a probability reads
\bea
\cP(x(t_1)&=&Sx_m,x(t_2)=Sx_{m-1},\dots,x(t_m)=Sx_1) =\\ &=&\cN_{\widehat{\cC}'}(x_1,x_2,\dots,x_m)
\eea
with
\bea
\widehat{\cC}'=
\begin{pmatrix}
S\cC(0)S & S\cC(t_2-t_1)S & \dots & S\cC(t_m-t_1)S \\
S\cC(t_1-t_2)S & S\cC(0)S & \dots & S\cC(t_m-t_2)S \\
\vdots & \vdots & \ddots & \vdots \\
S\cC(t_1-t_m)S & S\cC(t_2-t_m)S & \dots & S\cC(0)S \\
\end{pmatrix} 
\eea
The two probabilities are equals if and only if $\widehat{\cC}=\widehat{\cC}'$. For the validity of such a condition for whatever choice of observation times, one needs $S \cC(t) S = \cC(-t) = \cC(t)^T$ or, in Fourier space $S\cCfo S=\cCfo^T$. This is analogous to the renowned Onsager reciprocity relation~\cite{marconi2008fluctuation}. Actually the closest equivalent to original Onsager reciprocity is obtained by taking the time derivative of such relation and computing it in $t=0$, i.e. $SLS=L^T$, where $L=\dot\cC(0)$ is the Onsager matrix (see Chapter 5.3 of Gardiner's book~\cite{gardiner2009stochastic}).

We can also compute the Kullback–Leibler divergence $\cD_m$  between the above forward and reverse probabilities of $m$-paths, in order to mimic entropy production features (note, the true entropy production rate is typically computed for the continuous path probabilities, i.e. taking infinite observations at infinitesimal time-delays):
\bea
\cD_m &=& \int \prod_k dx_k \; \cN_{\widehat{\cC}}(x_1,...,x_m) \log{\frac{\cN_{\widehat{\cC}}(x_1,...,x_m)}{\cN_{\widehat{\cC}'}(x_1,...,x_m)}} = \\
&=&\frac{1}{2}\text{Tr}\lx(\widehat{\cC}\widehat{\cC'}^{-1} - I\rx)
\eea
From the above considerations, a main thing is evident: for 1-dimensional systems it is immediate to verify that, since $\cC(t)=\cC(-t)$ we have $\cD_m=0$, i.e. all groups of $m$ observations have identical forward and backward probabilities. This result is true whatever are the operators in the original equation, i.e. $\forall \cL,\cB\text{ and }\nu$.
\subsection{A Markovian case}
Equations like $\cL x=\cB\xi$ can arise from a genuine multidimensional Ornstein–Uhlenbeck process once we just observe a single dynamic variable or a linear combination of these.
Let us consider the following stochastic process in $D$ dimensions
\begin{eqnarray}\label{eq:OU}
\frac{d x}{dt} + A x = B\xi \qquad \ave{\xi_i(t)\xi_j(t')} = \nu_{ij}\delta(t-t'),
\end{eqnarray}
where $A$ is an invertible and positive definite $D \times D$ real matrix and $\nu$ is the covariance matrix of the noise.
It is convenient to rewrite the equation in the reference frame that has the eigenstates of symmetric matrix $B\nu B^T$ as a basis.
In this way we can interpret the contribution of noise in terms of something analogous to the temperatures $T_i$ of $D$ thermal bath, i.e.
\bea
 B\nu B^T= U \Sigma U^T \qquad UU^T=U^TU=I \qquad \Sigma_{ij} = 2T_i \delta_{ij}
\eea
so, by performing the substitutions 
$U^T x \to x$, $U^T A U \to A$ and $U^T B \xi \to \xi$
we obtain the effective process
\bea
\frac{d x}{dt} + A x = \xi \qquad \ave{\xi_i(t)\xi_j(t')} = 2T_i \delta_{ij}\delta(t-t'),
\eea
from which we verify that the statistical independence between the effective thermal baths $\ave{\xi_i \xi_j} \sim \delta_{ij}$ is not an approximation.
By direct integration we obtain the following expressions for response $\cR(t)$ and time correlation matrix $\cC(t)$ 
\begin{eqnarray}\label{acca}
\cR(t) &=& \theta(t)e^{-t A} \qquad\qquad \lx(\cR^\dagger(t) =\cR(-t)^T=\theta(-t)e^{t A^T}\rx) \nonumber \\
\cC(t) &=& \cR(t) C + C \cR(-t)^T \qquad \lx(C=\cC(0)\rx) \\ 
\dot{\cC}(t) &=& C A^T \cR(-t)^T - \cR(t) A C \\
\Sigma &=& AC+C A^T
\end{eqnarray}
Since the process is Markovian (and given the positivity of $A$) it has an invariant measure and it is easy to verify that the single-time normal distribution $\cN_C(x)$ is the stationary solution of the following Fokker-Plank equation
\bea
\frac{\partial}{\partial x}\lx( Ax + \frac{1}{2}\Sigma\frac{\partial}{\partial x}\rx) \cN_C(x)
= \frac{1}{2} \text{Tr}(2A-\Sigma C^{-1}) + \\
+\frac{1}{2}\sum_{ij} \lx(C^{-1}x\rx)_i\lx(\Sigma - 2AC\rx)_{ij}\lx(C^{-1}x\rx)_j = 0
\eea
because from (\ref{acca}) we have
\bea
(\Sigma - 2AC)^T &=& \Sigma -2CA^T = 2AC -\Sigma = -(\Sigma-2AC) \\ &\rightarrow&  \sum_{ij}y_i (\Sigma - 2AC)_{ij}y_j = 0 \quad\forall y \\
\text{Tr}(2A-\Sigma C^{-1}) &=& \text{Tr}(A - CA^TC^{-1}) = \\ 
&=& \text{Tr}(A)-\text{Tr}(A^T) = 0.
\eea
In Fourier space it is convenient to look at the following expression obtained by rationalizing the denominators of $\cC(\om)$ 
\bea
\cCfo &=& \lx(\om^2+A^2\rx)^{-1} \lx(\om^2 \Sigma + i\om(A\Sigma-\Sigma A^T) + A\Sigma A^T\rx) \times \\
&\times& \lx(\om^2+(A^T)^2\rx)^{-1} = \cCfo^\dagger \qquad 
\eea
so, by comparing it with its transpose
\bea
\cCfo^T &=& \lx(\om^2+A^2\rx)^{-1} \lx(\om^2 \Sigma - i\om(A\Sigma-\Sigma A^T) + A\Sigma A^T\rx) \times \\
&\times& \lx(\om^2+(A^T)^2\rx)^{-1}
\eea
we obtain that the equilibrium condition $S\cCfo S=\cCfo^T$ for $S=I$ (or $S=-I$) is $A\Sigma=\Sigma A^T$, ($A_{ij}T_j=A_{ji}T_i$). This condition slightly differs from the usual Onsager's relation $AC=CA^T$ but the two formulations are equivalent. Indeed,
since $C=\int_{0}^{+\infty}{\rm d}t \,e^{-tA}\,\Sigma\, e^{-t A^T}$
is the solution of $AC+CA^T=\Sigma$, $AC=CA^T$ reads
\bea
&&A\int_{0}^{+\infty}{\rm d}t \,e^{-tA}\,\Sigma\, e^{-tA^T}=\int_{0}^{+\infty}{\rm d}t \,e^{-tA}\,\Sigma\, e^{-tA^T}A^T \\ &&\implies
\int_{0}^{+\infty}{\rm d}t \,e^{-tA}\,\left[A\Sigma-\Sigma A^T\right] e^{-tA^T} = 0 \\
&&\implies A\Sigma=\Sigma A^T
\eea
Furthermore, if $A\Sigma=\Sigma A^T$ we have
\bea
AC&=&\int_{0}^{+\infty}{\rm d}t \,e^{-tA}\,A\Sigma\, e^{-tA^T}=\\
&=&\int_{0}^{+\infty}{\rm d}t \,e^{-tA}\Sigma A^T\, e^{-tA^T}=CA^T.
\eea
Hence, $AC=CA^T\iff A\Sigma=\Sigma A^T$. Interestingly, if the structure of $A$ is not appropriate, the equilibrium with $D$ effective thermal baths is impossible. In fact, by assuming $A_{ij} \neq 0 \; \forall i<j$, we must have
\bea
\lx\{
\begin{array}{rcl}
   T_j &=& T_i A_{ji}/A_{ij}  \\
   T_k &=& T_j A_{kj}/A_{jk} = T_i A_{ji}A_{kj}/A_{ij}A_{jk} = T_i A_{ki}/A_{ik}\\
\end{array}\rx.
\eea
This implies $A_{ij}A_{jk}A_{ki}=A_{ik}A_{kj}A_{ji} \; \forall i<j<k$.
On the other hand, the condition equilibrium implies pure imaginary poles only in $\cCfo$. In fact, if $T_i>0 \; \forall i$, since $A\Sigma=\Sigma A^T$, we can diagonalize the symmetric matrix $\widehat{A} = \Sigma^{-\frac{1}{2}} A \Sigma^{\frac{1}{2}}=\widehat{A}^T = U \Lambda U^T$ in order to derive the eigenstates of $A=\Sigma^{\frac{1}{2}}U\Lambda U^T \Sigma^{-\frac{1}{2}} = V\Lambda V^{-1}$ and verify, due to the spectral theorem for symmetric matrices, that all its eigenvalues are real and then $\cC(t)$ is a sum of pure real exponential. In the case there are some $T_i=0$ we simply separate the two types of variables and we look at the blocks of the matrices
\bea
\Sigma = 
\begin{pmatrix}
0 & 0 \\
0 & \Sigma' \\
\end{pmatrix}
\qquad \Sigma'_{ij}=2T_i\delta_{ij}
\qquad
A = 
\begin{pmatrix}
\alpha & -\lambda \\
-\mu & \gamma \\
\end{pmatrix}
\eea
In this way, the condition $A\Sigma=\Sigma A^T$ implies $\lambda=0$ and real eigenvalue for $\gamma$ since $\gamma\Sigma'=\Sigma'\gamma^T$ and then
\bea
\cCfo = 
\begin{pmatrix}
0 & 0 \\ 
0 & (i\om + \gamma)^{-1}\Sigma' (-i\om+\gamma^T)^{-1} \\
\end{pmatrix}
\eea
which has pure imaginary poles only.\\
Note that the equilibrium condition $A\Sigma=\Sigma A^T$ implies also the familiar fluctuation-response theorems for equilibrium systems i.e. ($\theta(0)=\frac{1}{2}$)
\bea
\cC(t) = (\cR(t) + \cR(-t) ) C &\to&  \cCfo = 2\text{Re} \cRfo C \\
\dot{\cC}(t) = (\cR(-t) -\cR(t)) \Sigma &\to&  \cCfo = -\frac{1}{\om} \text{Im} \cRfo \Sigma
\eea
If some dynamical variables change the sign in the reverse trajectory ($S\neq \pm I$) the generalization it is quite more complicated. To begin, we move all the dynamic variables that change sign at the bottom of the vector and we look at all the matrices as made up of 4 blocks
\bea
&&M=\begin{pmatrix} M_{++} & M_{+-} \\ M_{-+} & M_{--}\end{pmatrix}
\qquad S=\begin{pmatrix} I & 0 \\ 0 & -I \end{pmatrix} \\
&&SMS=\begin{pmatrix} M_{++} & -M_{+-} \\ -M_{-+} & M_{--}\end{pmatrix}
\eea
In this way equilibrium condition $S\cCfo S=\cCfo^T$ simply reads $\text{Re}(\cCfo_{+-}) = 0$ where (I omit the $\om$ dependence) 
\bea
\cCf_{+-} &=& \cRf_{++}\cSgf_{++}\cRf_{-+}^\dagger +\cRf_{++}\cSgf_{+-}\cRf_{--}^\dagger +\\ &+&\cRf_{+-}\cSgf_{-+}\cRf_{-+}^\dagger +\cRf_{+-}\cSgf_{--}\cRf_{--}^\dagger
\eea
The computation is considerably simplified in the case of a "symplectic" stochastic dynamics like
\bea
\lx\{
\begin{array}{l}
    \dot{x} = y \\
    M\dot{y} = -\Gamma y - K x + B \xi  \\
\end{array}\rx.
\quad  \ave{\xi_i(t)\xi_j(t')} = \nu_{ij} \delta(t-t') \\
 \lx( x,y\in\mathbb{R}^D \quad M,\Gamma,K,B,\nu \in \mathbb{R}^{D \times D} \rx)
\eea
for which we can forget that $x$ and $y$ have opposite time reversal parity simply by considering the second order stochastic equation
\bea
M\ddot{x} +\Gamma \dot{x} + K x = B \xi 
\eea
and the time correlation function $\cC_{ij}(t-t')=\ave{x_i(t)x_j(t')}$ which involves just the $x$ components.
In fact, if $M$ is invertible, we are authorized to simplify thanks to the following substitutions
\bea
&&U^T M x \to x \qquad U^T\Gamma M^{-1} U \to \Gamma \\
&& U^TK M^{-1} U \to K \qquad U^T B \xi \to \xi
\eea
where, as in the general case, we are put ourself in the reference frame that diagonalize $B\nu B^T=U\Sigma U^T$.\\
In this way we get
\bea
\ddot{x} +\Gamma \dot{x} + K x = \xi \qquad \ave{\xi_i(t)\xi_j(t')} = 2T_i \delta_{ij} \delta(t-t')
\eea
from which 
\bea
\cCfo &=& \lx(-\om^2 +i\om \Gamma + K\rx)^{-1} \Sigma (-\om^2 -i\om\Gamma^T + K^T)^{-1}  \\
&=& (K-\om^2)^{-1}(I + i\om(K-\om^2)^{-1}\Gamma)^{-1}\Sigma \times \\
&\times& (I - i\om\Gamma^T(K^T-\om^2)^{-1})^{-1}(K^T-\om^2)^{-1}
\eea
($\Sigma_{ij} = 2T_i \delta_{ij}$)
As in the general case, by rationalizing and by imposing $\cCfo=\cCfo^T$ we are able to prove that the equilibrium holds when both the conditions $\Gamma \Sigma = \Sigma \Gamma^T$ ($\Gamma_{ij}T_j=\Gamma_{ji}T_i$) and $K \Sigma \Gamma^T= \Gamma \Sigma K^T$ are simultaneously satisfied.
Even in this case we have the further condition $\Gamma_{ij}\Gamma_{jk}\Gamma_{ki}=\Gamma_{ik}\Gamma_{kj}\Gamma_{ji}$ without which equilibrium with $D$ thermal bath is impossible and, by following the same procedure as in the general case,  $\Gamma$ has real eigenvalues only.
It is possible to prove that $K$ also has real eigenvalues: it is sufficient to make explicit the eigenstates of $\Gamma=\Sigma^{\frac{1}{2}}U\Lambda U^T \Sigma^{-\frac{1}{2}}$ and to note that the condition $K\Sigma\Gamma^T=\Gamma\Sigma K^T$ implies that the matrix $\widehat{K}=\Lambda^{-\frac{1}{2}}U^T\Sigma^{-\frac{1}{2}}K\Sigma^{\frac{1}{2}}U\Lambda^{\frac{1}{2}}$ must be symmetric from which immediately we deduce that the eigenvalues of $K$ are real too.
\subsection{One variable from a Markovian system}
We look now at a single scalar component $y$ of a generic multidimensional Ornstein-Uhlenbeck process.
We showed that, unless pathological situations, the linearity of the equation allows us to scale suitably the dynamical variables and to look at the stochastic process in the reference frame for which the covariance matrix of the noise is diagonal and the coefficient in front at the higher order derivative is unitary.
By focusing on this "effective" process, we minimized the number of involved parameters and simplified the calculations relating to the equilibrium conditions but we have lost the identity as a component of the state vector of the original dynamical variable $y$ which we are observing. However, surely $y$ will be a linear combination of the components of the effective process so, we can still derive some general considerations about the properties of $y$ just by looking at the time correlation function $\cC_y(t)$ of a generic linear combination $y=\sum_i a_i x_i$ of the roto-scaled dynamical variables $x_i$, i.e.
\bea
\cC_y(t-t') &=& \ave{y(t)y(t')}_c = \sum_{ij} a_i \cC_{ij}(t-t') a_j \\
\widetilde{\cC}_y(\om) &=& 2\sum_i T_i \sum_{jk} a_j (i\om+A)^{-1}_{ji} (-i\om+A^T)^{-1}_{ik} a_k \\
&=& 2\sum_i T_i \lx|\sum_j a_j (i\om+A)^{-1}_{ji}\rx|^2
\eea
If we imagine to perform the inverse $(i\om+A)^{-1}$ with the Cramer's rule for which $M^{-1} = \text{adj}(M)/\text{det}(M)$ where $\text{adj}(M)$ is the transpose of cofactor of $M$ we get
\begin{eqnarray}\label{co}
\widetilde{\cC}_y(\om) &=&
2\frac{\sum_i T_i \lx|\widetilde{\cB}_i(\om)\rx|^2}{\lx|\cLfo\rx|^2} = \sum_i  T_i f_i(\om) \\ f_i(\om)&=&2\frac{\lx|\widetilde{\cB}_i(\om)\rx|^2}{\lx|\cLfo\rx|^2} \nonumber
\end{eqnarray}
where $\cLfo$ and $\widetilde{\cB}_i(\om)$ are $i\om$-polynomials with real coefficients that we can write by looking at their root $-\lambda_\alpha$ and $-\gamma_{i\beta}$ respectively
\begin{eqnarray*}
\cLfo&=&\text{det}(i\om+A) = \prod_{\alpha=1}^D (i\om + \lambda_\alpha) \\
\widetilde{\cB}_i(\om)&=&\sum_j a_j \text{adj}(i\om+A)_{ji} = \cB_i \prod_{\beta=1}^{D-1} (i\om + \gamma_{i\beta}) \\
f_i(\om) &=&  2\lx|\cB_i\rx|^2\frac{\prod_{\beta=1}^{D-1}\lx|\om-i\gamma_{i\beta}\rx|^2}{\prod_{\alpha=1}^D \lx|\om -i\lambda_\alpha\rx|^2}
\end{eqnarray*}
In this way it is evident, as expected, that a stochastic equations satisfied by $y$ can be formally written as
\bea
\cL y = \sum_i \cB_i \xi_i \qquad \ave{\xi_i(t)\xi_j(t')} = 2T_i \delta_{ij}\delta(t-t').
\eea
which is in the form of the equation (\ref{genlinequ}).
Note that the equation satisfied by the $n$-th derivative of $y$,
$z=\partial^n y/\partial t^n$, since $\widetilde{z}(\om)=(i\om)^n\widetilde{y}(\om)$ in Fourier space simple reads 
\bea
\cLfo \widetilde{z}(\om)=(i\om)^n\sum_i \cBf_i(\om)\xif_i(\om)
\eea
from which we immediately derive that $\cCf_{z}(\om)=\om^{2n} \cCf_y(\om)$.
By using the residue theorem and the complex conjugate root theorem which states that the non-real roots of a polynomial with real coefficient appear always into pairs of complex conjugates (this implies $f_i(-z)=f_i(z)$ and $f_i(\overline{z})=\overline{f_i(z)}$ $\forall z\in \mathbb{C}$) we are able to compute $\cC(t)$. First of all, the functions $f_i(\om)$ have the same denominator then the same $2D$ singularities in the complex plane so,
if $\text{Im}\lambda_\alpha=0$ we have two pure imaginary poles in $i\lambda_\alpha$ and $-i\lambda_\alpha$ while
if $\text{Im}\lambda_\alpha \neq 0$ we have four complex poles in 
$\pm i\lambda_\alpha$ and $\pm i \overline{\lambda_\alpha}$.
We indicate with $\mu_\alpha$ the set of $D$ poles that have a positive imaginary part then, for $t>0$, we get
\begin{eqnarray}\label{Ct}
\cC_y(t)&=&\int \frac{d\om}{\sqrt{2\pi}}\,\widetilde{\cC}_y(\om) e^{i\om t} = \sum_{\alpha=1}^D c_\alpha e^{i\mu_\alpha t} \nonumber \\
c_\alpha &=& \sqrt{2\pi}i\sum_{i} T_i \text{Res}(f_i(\om),\mu_\alpha)
\end{eqnarray}
Finally, the properties of the functions $f_i(\om)$ and the simultaneous presence of the poles $\mu_\alpha$ and $-\overline{\mu_\alpha}$ implies
\begin{eqnarray*}
    \cC_y(t) &=&\underset{\text{Im}\lambda_\alpha=0}{\sum_{\alpha:}} c_\alpha^{(0)} e^{-\lambda_\alpha t} +  \qquad (\forall t > 0) \\ &+&\underset{\text{Im}\lambda_\alpha > 0}{\sum_{\alpha:}} e^{-\text{Re}\lambda_\alpha t} \lx( c_\alpha^{(+)} \cos{\text{Im}\lambda_\alpha t} + c_\alpha^{(-)} \sin{\text{Im}\lambda_\alpha t} \rx) 
\end{eqnarray*}
where the $c_\alpha^{(\star)}$ are a total of $D$ real constant which can be expresses as a linear combinations of temperatures $T_i$,
i.e. $c_\alpha^{(\star)} = \sum_i T_i c_{i\alpha}^{(\star)}$. If the equilibrium condition $A_{ij}T_j=A_{ji}T_i$ holds (or $\Gamma_{ij}T_j=\Gamma_{ji}T_i$ in the symplectic case), it means that we can write all temperatures $T_i$ as a function of only one, for example, $T_1=T$ and $T_i=T A_{i1}/A_{1i}(=T\Gamma_{i1}/\Gamma_{1i})= T g_i$ and then
\bea
c_\alpha^{(\star)} = T \sum_i g_i c_{i\alpha}^{(*)} = T d_\alpha^{(\star)}
\eea
where the $d_\alpha^{(\star)}$ depend on the drift only. In other word: equilibrium implies a single effective thermal bath.

\section*{Appendix C}\label{appendix:Integration_Scheme}
To perform a numerical integration of the equations of motion, one could use one of the standard algorithms for stochastic differential equations such as the Euler or Runge-Kutta stochastic method, just to give two examples. However, these methods require the use of integration time steps much smaller than the characteristic times of the system. In the case of the Ornstein-Ulhenbeck process, however, it is possible to use an exact algorithm.
We consider the following Cauchy problem
\begin{equation*}
\lx\{
\begin{array}{l}
\dot{x}+Ax=B\xi \\
x(t_0)=x_0 \\ 
\end{array}\rx. \qquad \ave{\xi_i(t)\xi_j(t')}=\nu_{ij}\delta(t-t')
\end{equation*}
for which the formal solution after a time-step $\eps$ is 
\bea
x(t_0+\eps) &=& e^{-\eps A}\lx( x(t_0) + \int_0^{\eps} ds \, e^{ s A} \xi(s) \rx) = \\
&=& e^{-\eps A}\lx( x(t_0) + w^{(\eps)} \rx) = R^{(\eps)} (x(t_0)+w^{(\eps)} )
\eea
The $D$ dimensional random vector $w^{(\eps)}$ is normally distribuited with average $\ave{w^{(\eps)}_i}=0$ and covariance matrix $C^{(\eps)}_{ij}=\ave{w^{(\eps)}_i w^{(\eps)}_j}$ equal to
\bea
C^{(\eps)} &=& \int_0^{\Delta t} ds \, e^{s A}  B \nu B^T e^{s A^T} = \\
&=&V \lx( \int_0^{\Delta t} ds\, e^{s \Lambda } \lx( V^{-1} B \nu B^T V^{-T} \rx) e^{s \Lambda} \rx) V^T = \\
&=&V \lx( \int_0^{\Delta t} ds\, e^{s \Lambda } \Sigma e^{s \Lambda} \rx) V^T = V H^{(\eps)} V^T
\eea
where $\Lambda$ and $V$ diagonalize $A=V\Lambda V^{-1}$ ($\Lambda_{ij}=\lambda_i \delta_{ij}$), $\Sigma=V^{-1} B \nu B^T V^{-T}=\Sigma^T$ and
\bea
H^{(\eps)} &=& \int_0^{\Delta t} ds\, e^{s \Lambda } \Sigma e^{s \Lambda} =H^T \\
H^{(\eps)}_{ij} &=& \Sigma_{ij} \frac{e^{\eps(\lambda_i+\lambda_j)} - 1}{\lambda_i + \lambda_j} = H^{(\eps)}_{ji}
\eea
In general $H^{(\eps)}$ is a complex matrix but the spectral theorem for symmetric matrix assures us that it can still be decomposed into $H^{(\eps)}= U M^{(\eps)} U^T$ where $U$ is an unitary matrix and $M^{(\eps)}_{ij}=\sigma_i^2 \delta_{ij}$ is a real diagonal matrix with non-negative entries.
This suggests that, once $\Delta t$ is fixed, matrix $A$ and $H^{(\eps)}$ are diagonalized and $R^{(\eps)}=Ve^{-\eps \Lambda}V^{-1}$ is computed, it is possible to introduce an exact integration algorithm by implementing the following instructions step by step
\begin{enumerate}
    \item sample $D$ random numbers $z_i$ from a normal distributions of zero mean and relative variances $\sigma_i^2$
    \item compute the vector $w^{(\eps)}= V U z$
    \item iterate with $x(t+\Delta t) = R^{(\eps)} \lx(x(t) + w^{(\eps)}\rx)$.
\end{enumerate}
\bibliographystyle{unsrt}
\bibliography{biblio}
\end{document}